\documentclass{article}

\usepackage[dvips]{color}

\usepackage{amssymb}

\newcommand{\beq}[1]{\begin{equation} \label{#1} }
\newcommand{\beqa}[1]{\begin{eqnarray} \label{#1} }
\newcommand{\eeq}{\end{equation}}
\newcommand{\eeqa}{\end{eqnarray}}
\newcommand{\ba}[1]{\begin{array}{#1}}
\newcommand{\ea}{\end{array}}
\newcommand{\rf}[1]{(\ref{#1})}
\newcommand{\bosy}[1]{ \mbox{\boldmath ${#1}$} }

\newcommand{\hbC}{\widehat{\bf C}}  	\newcommand{\hbS}{\widehat{\bf S}}
  				\newcommand{\hS}{\widehat{S}}
\newcommand{\hbI}{\widehat{\bf I}} 
\newcommand{\hbP}{\widehat{\bf P}}  	\newcommand{\hbQ}{\widehat{\bf Q}}
\newcommand{\cC}{{\mathbb C}} 				\newcommand{\cS}{{\mathbb S}}

\begin{document}

\title{ \textcolor{blue}{Optimal Orientation of Anisotropic Solids} }

\author{A. N. Norris\\ \\    Mechanical and Aerospace Engineering, 
	Rutgers University, \\ Piscataway NJ 08854-8058, USA norris@rutgers.edu } 
	\date{}
\maketitle

\begin{abstract} 
Results are presented for finding the optimal orientation of an anisotropic elastic material.  The problem is formulated as minimizing the strain energy subject to rotation of the  material axes, under a state of uniform stress.  It is shown that  a stationary value of the strain energy requires the stress and strain tensors to have a common set of principal axes.  The new derivation of this well known coaxiality condition uses the 6-dimensional expression of the rotation tensor for the elastic moduli.  Using this representation it is shown that  
the stationary condition is a minimum or a maximum if an explicit set of conditions is satisfied.  Specific  results are given for materials of cubic, transversely isotropic (TI) and tetragonal symmetries. In each case the existence of a minimum or maximum depends on the sign of a single elastic constant.  The stationary (minimum or maximum) value of energy can always be achieved for cubic materials.  Typically, the optimal orientation of a solid with cubic material symmetry is not aligned with the symmetry directions.  Expressions are given  for the optimal orientation of  TI and tetragonal materials,   and  are in agreement with results of  Rovati and Taliercio \cite{Rovati03} obtained by a different procedure.  A new concept is introduced,  the strain deviation angle, which defines the degree to which a state of stress or strain is not optimal.  The strain deviation angle is zero for coaxial stress and strain.  An approximate formula is given for the strain deviation angle  which is valid for materials that are weakly anisotropic. 
\end{abstract}

\section{Introduction}

The strain energy of a piece of homogeneous anisotropic elastic material depends on the orientation of the material relative to the directions of principal stress, although the orientation dependence vanishes trivially for isotropic solids.  This property is therefore an inherently anisotropic feature of elasticity, and it raises the question of how to find  the material orientation (if any) which minimizes the strain energy for a given state of stress or  strain. New results are presented in this paper on the determination of optimal orientations for both general and specific types of anisotropy. 

The general  problem of determining optimal orientations in anisotropic elasticity has been the subject of several studies in the last two decades, beginning with the work of Seregin and Troitskii \cite{Seregin81} in the context of  orthotropic solids. They determined the important {\it coaxiality condition}: a minimum or maximum of strain energy requires that the stress and strain share common principal axes.  The coaxiality condition was subsequently and independently obtained by others: first, by  Rovati and Taliercio \cite{Rovati91} who considered 3D  elastic materials with orthotropic and cubic symmetries (although their derivation is not restricted to these symmetries but is applicable to general anisotropy), and later by    
Cowin \cite{Cowin94}.  Cowin  derived the coaxiality condition  independent of material symmetry considerations.  He showed that the commutativity of the stress and strain is a consequence of the stationarity condition of the strain energy with respect to rotations of the moduli.  Vianello \cite{Vianello96a} provided a more formal derivation of coaxiality in linear elasticity.  He used the  tangent space of the rotation group to show that there are at least two orientations of the moduli that simultaneously make the energy stationary and stress and strain coaxial, a result later generalized to  hyperelasticity \cite{Vianello96b} (it was subsequently shown that at least three such orientations exist for both linear elasticity \cite{Sgarra97a} and for hyperelasticity \cite{Sgarra97b}).  There is a slight distinction between the problems considered by Cowin and by Vianello that is important to note for our purposes \cite{Cowin97}.  Thus, Cowin \cite{Cowin94} considered  stress states with fixed principal directions but arbitrary amplitudes, whereas Vianello \cite{Vianello96a} assumed a specific state of stress.  The former constraint  defines a smaller set of possible elastic moduli for which coaxility can be attained, because it requires that optimal condition be simultaneously satisfied by a family of coaxial stresses.  Not surprisingly, Cowin found that only materials with orthotropic symmetry meet these conditions.   In this paper the stress state is taken as given, in the same spirit as \cite{Vianello96a} and also of Rovati and Taliercio \cite{Rovati03}.   While the emphasis here is on three dimensional elasticity, the optimality problem has also been 
 addressed within  the context of  two dimensional elasticity \cite{Pedersen89,Pedersen90,Gea04}.
Cowin and Yang  \cite{Cowin00} considered a related but more general question of optimality with respect to Kelvin modes, rather than simply the freedom to orient a given material.  
  A more extensive review of the literature can be found in the recent paper of Rovati and Taliercio \cite{Rovati03}.  

It is interesting to note that the coaxiality condition has been derived in a variety of  different ways: for particular symmetries (e.g. orthotropic) \cite{Seregin81}, using Lagrange multipliers \cite{Cowin94}, from general analytic considerations \cite{Vianello96a}, and even using the 6-dimensional eigenvector properties of the elasticity tensor \cite{Banichuk96}. The derivation of the coaxiality condition  presented here differs from all these previous methods.    Our starting point is a representation of the rotation matrix due to Mehrabadi, Cowin and Jaric \cite{mcj}.  This formulation also enables derivation of conditions for minima or maxima, in a simpler  and more general form than that obtained by Cowin \cite{Cowin94}.  Section \ref{sec2} begins with the problem definition and notation.  The stationarity and min/max conditions are discussed in Section \ref{sec3}.  Specific conditions for particular material symmetries are derived in Section \ref{sec4}, and we conclude in Section \ref{sec5} by defining the {\it strain deviation angle}, a concept which could have application in practical circumstances in anisotropic elasticity.

\section{Problem definition and notation}\label{sec2}

\subsection{Optimal orientation of anisotropic solids}

Consider a fixed coordinate system $\{ \bf{e}_1,\, \bf{e}_2,\, \bf{e}_3 \}$ coincident with the principal axes of stress.  The stress tensor is therefore  
\beq{eq1}
\bosy{\sigma} = \sigma_{\rm I}  \,\bf{e}_1 \otimes  \bf{e}_1 + \sigma_{\rm II} \, \bf{e}_2 \otimes  \bf{e}_2 + \sigma_{\rm III} \, \bf{e}_3 \otimes  \bf{e}_3, 
\eeq
where $\sigma_{\rm I}$, $\sigma_{\rm II}$ and $\sigma_{\rm III}$ are the principal stresses, in no particular order.  Alternative expressions for the stress include the $3\times 3$ matrix representation, 
\beqa{s1}
 \bosy{\sigma}    = 
\left[ \begin{array}{ccc}
\sigma_{\rm I} & 0 & 0 \\
0 & \sigma_{\rm II} &  0 \\
0 & 0 &  \sigma_{\rm III}  
\end{array} \right] , 
\eeqa 
and indicial\footnote{Lower case Latin suffices take on the values 1, 2, and 3, and the summation convention on repeated indices is assumed unless noted otherwise.} notation, 
\beq{t2}
\sigma_{ij}=  \sigma_{\rm I}\,  \delta_{i1} \delta_{j1} +  \sigma_{\rm II}\,  \delta_{i2} \delta_{j2}+  \sigma_{\rm III}\,  \delta_{i3} \delta_{j3}. \quad
\eeq
Our goal is to find the orientation or orientations which minimize the energy function for fixed stress
\beq{s4}
{\cal E}  \equiv  \sigma_{ij} \sigma_{kl} \, s_{ijkl}\, . 
\eeq 
Here, $s_{ijkl}$ are the components of the fourth order compliance tensor relative to $\{ \bf{e}_1,\, \bf{e}_2,\, \bf{e}_3 \}$.  Think of the material as being free to be oriented in  such a way that the energy ${\cal E}$ depends upon the orientation of the {\it material}  with respect to the fixed principal axes of the stress. 
The material moduli for stiffness and compliance are $\cC^{(0)}$ and $\cS^{(0)}$ when aligned with the fixed axes. It is not necessary  to  specify at this stage whether or not the moduli possess any symmetry with respect to these axes.  The main point is that  the material is free to orient in arbitrary directions with {\it oriented} moduli  $\cC$ and $\cS$ while the stress  orientation remains  fixed. 

\subsection{Notation and tensor rotation}
 Hooke's law  relating stress $\sigma_{ij}$ and strain $\varepsilon_{ij}$ is 
\beq{00}
\sigma_{ij} = c_{ijkl}\varepsilon_{kl}, 
\qquad
\varepsilon_{ij} = s_{ijkl}\sigma_{kl}. 
\eeq
Here $c_{ijkl}$ denote the components of the stiffness tensor, inverse to the compliance: 
$c_{ijkl}s_{klpq} = I_{ijpq}$, where $I_{ijpq} = (\delta_{ip} \delta_{jq} +\delta_{iq}\delta_{jp})/2$ is the fourth order identity tensor.   
The rotated elasticity components could be expressed in terms of the unrotated components $c^{(0)}_{ijkl}$ and $s^{(0)}_{ijkl}$, using Euler angles, for instance.  The concise Voigt notation is used to represent the elements of the elasticity tensor in the fixed basis.     Thus, the compliance is  
$\cS = \left[ S_{IJ} \right],\  I,J=1,2,\ldots 6$, with $S_{12}= s_{1122}$, $S_{16}= s_{1112}$,  $S_{44}= s_{2323}$, etc.,   
\beqa{a1a}
\cS  = \left[ 
\ba{cccccc}
S_{11} & S_{12} & S_{13} & 
 S_{14} &  S_{15} & S_{16} 
\\ & & & & & \\
 & S_{22} & S_{23} & 
 S_{24} &  S_{25} &  S_{26} 
\\ & & & & & \\
 & & S_{33} & 
  S_{34} &  S_{35} &  S_{36} 
\\ & & & & & \\
 &   &   
 & S_{44} & S_{45} & S_{46}
\\ & & & & & \\
   S &Y  &M & & S_{55} & S_{56}
\\ & & & & & \\
&&&& & S_{66}
\ea \right] .  
\eeqa

An alternative representation for the elasticity tensor, closely related to \rf{a1a}, is the $6\times 6$ matrix   $\hbS$ with elements $ \left[ \hS_{IJ}\right]$  defined as 
\beqa{defs}
\hbS = {\bf T} \, \cS\, {\bf T},
\quad \mbox{where  }
{\bf T} \equiv 
\left[ 
\ba{cc}
~ {\bf I}~ & 0  \\ & \\
0 & \sqrt{2} {\bf I}
\ea \right] .
\eeqa
Explicitly,   
\beqa{a1}
\hbS  \equiv \left[ 
\ba{cccccc}
S_{11} & S_{12} & S_{13} & 
 2^{\frac12} S_{14} & 2^{\frac12} S_{15} & 2^{\frac12} S_{16} 
\\ & & & & & \\
S_{12} & S_{22} & S_{23} & 
 2^{\frac12} S_{24} & 2^{\frac12} S_{25} & 2^{\frac12} S_{26} 
\\ & & & & & \\
S_{13} & S_{23} & S_{33} & 
 2^{\frac12} S_{34} & 2^{\frac12} S_{35} & 2^{\frac12} S_{36} 
\\ & & & & & \\
 2^{\frac12} S_{14} & 2^{\frac12} S_{24} & 2^{\frac12} S_{34} 
 & 2S_{44} & 2S_{45} & 2S_{46}
\\ & & & & & \\
 2^{\frac12} S_{15} & 2^{\frac12} S_{25} & 2^{\frac12} S_{35} 
 & 2S_{45} & 2S_{55} & 2S_{56}
\\ & & & & & \\
 2^{\frac12} S_{16} & 2^{\frac12} S_{26} & 2^{\frac12} S_{36} 
 & 2S_{46} & 2S_{56} & 2S_{66}
\ea \right] , 
\eeqa
This representation is useful in taking advantage of the fact that  fourth order elasticity tensors in 3 dimensions are 
equivalent to second order symmetric tensor of 6 dimensions \cite{c3}.
Similar equations follow for $\cC$ and  $\hbC = {\bf T} \, \cC\, {\bf T}$.  Define 
\beqa{eqs2}
\widehat{\bosy{\sigma}}  \equiv \left[ 
\ba{c}
\sigma_{11} \\  \sigma_{22} \\ \sigma_{33} \\ \sqrt{2} \sigma_{23} \\  \sqrt{2}\sigma_{31} \\ \sqrt{2}\sigma_{12}
 \ea \right] , 
 \qquad
 \widehat{\bosy{\varepsilon}}  \equiv \left[ 
\ba{c}
\varepsilon_{11} \\  \varepsilon_{22} \\ \varepsilon_{33} \\ \sqrt{2}\varepsilon_{23} \\  \sqrt{2}\varepsilon_{31} \\ \sqrt{2}\varepsilon_{12}
 \ea \right] , 
\eeqa
then the stress strain relations \rf{00} become 
\beqa{s3}
\widehat{\bosy{\sigma}}   = \hbC \widehat{\bosy{\varepsilon}} , 
 \qquad
 \widehat{\bosy{\varepsilon}}  = \hbS \widehat{\bosy{\sigma}}. 
\eeqa
Note that $\hbS$ and $\hbC$ are the matrix inverse of each other; 
$\hbS \hbC = \hbC \hbS = \hbI$ where $\hbI=$diag$(1,1,1,1,1,1)$. 

The rotation about ${\bf n}$, $|{\bf n} | = 1$  by angle $\phi$ is defined as ${\bf Q} ({\bf n} , \phi )\in O(3)$, such that vectors (including the basis vectors) transform as ${\bf v} \rightarrow {\bf v}' = {\bf Q}{\bf v}$. By considering small rotations, it may be readily seen that  ${\bf Q} ({\bf n} , \phi )$ can be expressed in terms a skew symmetric tensor ${\bf P}$ that is linear in $\bf n$. Thus,  
\beq{qsat1}
\frac{d {\bf Q}}{d\phi} ({\bf n} , \phi )  = {\bf P}\, {\bf Q} ({\bf n} , \phi ), \quad \mbox{where}\quad P_{ij}({\bf n})= e_{ijk}n_k\, ,  
\eeq
and hence 
\beq{qeq}
{\bf Q} = e^{\phi {\bf P}},  
\eeq
Note that    ${\bf Q}$ possesses alternate well-known expressions
\beqa{a1a1} {\bf Q}  ({\bf n} , \phi ) &=& {\bf I} + \sin\phi\, {\bf P}+ (1 -  \cos\phi) {\bf P}^2 \nonumber \\
&=& {\bf n}\otimes{\bf n} + \cos\phi \, ({\bf I} -{\bf n}\otimes{\bf n} )+
\sin\phi\, {\bf P}\, . 
\eeqa
In particular for our needs here, the small angle expansion is 
\beq{qsat2}
{\bf Q} ({\bf n} , \phi )  = {\bf I} +  \phi {\bf P} + \mbox{O}(\phi^2)\, .  
\eeq

Under the change of basis associated with ${\bf Q} ({\bf n} , \phi )$, second order tensors (including stress and strain) transform as $\bosy{\sigma} \rightarrow \bosy{\sigma} '$, where
\beq{qsat3}
\sigma_{ij} ' = Q_{ir}Q_{js} \, \sigma_{rs} 
\quad \Leftrightarrow \quad \sigma_{ij} ' = {\cal Q}_{ijrs}\, \sigma_{rs}. 
\eeq
The fourth order ``rotation" tensor follows from \rf{qsat3} as 
\beq{qsat4}
{\cal Q}_{ijrs} = \frac12 \, \big( Q_{ir}Q_{js} + Q_{is}Q_{jr}\big),  
\eeq
and eqs. \rf{qsat1} and \rf{qsat4} imply
\beq{qsat5}
\frac{d {\cal Q}_{ijrs} }{d\phi} ({\bf n} , \phi )  = {\cal P}_{ijpq}{\cal Q}_{pqrs}, 
\quad \mbox{with}\quad {\cal Q}_{ijrs}  ({\bf n} , 0 ) = I_{ijpq}, 
\eeq
where
\beq{qsat6}
{\cal P}_{ijpq} = 
\frac12 \, \big( \delta_{ip}P_{jq}  +\delta_{iq}P_{jp} +\delta_{jp}P_{iq} + \delta_{jq}P_{ip}
\big)\, . 
\eeq
The formal solution of \rf{qsat5}, with meaning that should be clear, is 
\beq{qsat7}
{\bf {\cal Q}} = e^{\phi {\bf {\cal P}}},   
\eeq
and the small angle approximation is
\beq{qsat8}
{\cal Q}_{ijpq} = I_{ijpq}+\phi  {\cal P}_{ijpq} + \mbox{O}(\phi^2)\, . 
\eeq

Mehrabadi et al. \cite{mcj}  derived an elegant expression for ${\bf {\cal Q}}$ analogous to the representation  for ${\bf Q}({\bf n} , \phi ) $. The key is the characteristic equation of ${\bf {\cal P}}$, (${\bf {\cal P}}^5 +5{\bf {\cal P}}^3+4{\bf {\cal P}}=0$ where ${\cal P}^2_{ijkl} = {\cal P}_{ijpq}{\cal P}_{pqkl}$ etc.) which permits the exponential expression \rf{qsat7} to be simplified.  The result is most simply presented in terms of the $6\times 6$ rotation matrix $\hbQ $  introduced by Mehrabadi et al. \cite{mcj}, and defined in the same manner as before.  Thus, 
$\hbQ  = {\bf T} {\cal Q} {\bf T} $  and $\hbP  = {\bf T} {\cal P} {\bf T} $  where ${\bf T}$ is defined in \rf{defs} and ${\cal Q}$ and ${\cal P}$ are the $6\times 6$ Voigt matrices.   Explicitly,  $\hbP$ is a skew symmetric six dimensional tensor  linear in  $\bf n$, 
\beqa{a1b}
\hbP ({\bf n}) =  \left[ 
\ba{cccccc}
0 & 0 & 0 & 
        0 & \sqrt{2} n_2 & -\sqrt{2} n_3  
\\ & & & & & \\
0 & 0 & 0 & 
        -\sqrt{2} n_1 &0 &  \sqrt{2} n_3
\\ & & & & & \\
0 & 0 & 0 & 
        \sqrt{2} n_1 &  -\sqrt{2} n_2 &0
\\ & & & & & \\
0 & \sqrt{2} n_1 & -\sqrt{2} n_1   & 0 & n_3 & -n_2
\\ & & & & & \\
-\sqrt{2} n_2 &0 &  \sqrt{2} n_2  & -n_3 & 0 & n_1
\\ & & & & & \\
\sqrt{2} n_3 &  -\sqrt{2} n_3 &0 &n_2 & -n_1 & 0 
\ea \right] . 
\eeqa
 $\hbQ  ({\bf n} , \phi )$ is  an orthogonal second order tensor of six 
dimensions, satisfying $\hbQ\hbQ^T = \hbQ^T\hbQ = \hbI$.  
Equation \rf{qsat7} becomes
\beq{they}    
 \hbQ ({\bf n} , \phi ) =  e^{\phi  \hbP({\bf n})} ,
 \eeq
 and  has the explicit expansion  \cite{mcj}
\beq{mc2}
\hbQ ({\bf n} , \phi )= \hbI + 
\sin\phi\, \hbP
+ (1 -  \cos\phi)\, \hbP^2 
+\frac13 \sin\phi (1 -  \cos\phi)\,( \hbP + \hbP^3)
+\frac16  (1 -  \cos\phi)^2\,( \hbP^2 + \hbP^4)\, . \quad
\eeq
Finally, we note that  fourth order tensors $\cC  $ transforms as $\hbC \rightarrow \hbC ' = \hbQ\hbC\hbQ^T$.

\subsection{Orientation function revisited}

Denote the matrix of rotation from the fixed to the ``current''  axes as $\hbQ $.  Thus, 
\beq{s3b} 
\hbC   = \hbQ \hbC^{(0)} \hbQ^T , 
\qquad
\hbS   = \hbQ \hbS^{(0)} \hbQ^T . 
\eeq
Hence the objective function of   \rf{s4} for the stress-based energy minimization becomes 
\beq{s5}
{\cal E} = \widehat{\bosy{\sigma}}^T\,  \hbS\,  \widehat{\bosy{\sigma}}\, . 
\eeq
This is the starting point in the next section to derive conditions necessary for a minimum.  It is important to emphasize the initial assumption that the stress is aligned with the fixed axes, \rf{s1}, or in terms of $\widehat{\bosy{\sigma}}$,     
\beqa{s1a}
\quad 
\widehat{\bosy{\sigma}}  \equiv \left[ 
\ba{c}
\sigma_{\rm I} \\  \sigma_{\rm II} \\ \sigma_{\rm III} \\ 0 \\ 0 \\ 0
 \ea \right] .
\eeqa 
This ensures that the energy varies as the material axes are rotated (if the stress were also rotated then the energy would be, trivially,  unchanged). 

\section{Stationarity and min/max conditions}\label{sec3}

\subsection{Angular derivatives of the strain energy}

Consider the energy  $\cal E$ of \rf{s5} as a function of the rotation $\hbQ$.  A  stationary value is obtained if $\cal E$ is unchanged  with respect to additional small rotations of $\hbS$.  This requires calculating the first derivative with respect to rotation angle for arbitrary rotation. The second derivative is needed  to distinguish the stationary point as a minimum or maximum, or otherwise. 

The strain energy obtained by arbitrary rotation  of the material about the axis $\bf n$ is 
\beq{s6}
{\cal E} ( {\bf n}, \phi) = \widehat{\bosy{\sigma}}^T\, \hbQ({\bf n} , \phi ) \hbS \hbQ^T ({\bf n} , \phi ) \,  \widehat{\bosy{\sigma}}\, .   
\eeq
The first derivative can be expressed, 
\beqa{s8q}
 \frac{\partial \, }{\partial \phi }{\cal E} ( {\bf n}, \phi) &= &
\widehat{\bosy{\sigma}}^T\, \big[ \hbP({\bf n}) \hbQ({\bf n} , \phi ) \hbS \hbQ^T ({\bf n} , \phi )  + \hbQ({\bf n} , \phi ) \hbS^T  \hbQ^T ({\bf n} , \phi )\hbP^T ({\bf n}) \big] \,  \widehat{\bosy{\sigma}}
\nonumber \\
&= & 2\widehat{\bosy{\sigma}}^T\,  \hbP({\bf n}) \hbQ({\bf n} , \phi ) \hbS \hbQ^T ({\bf n} , \phi )  \,  \widehat{\bosy{\sigma}}
\, .  
\eeqa
This follows from eqs. \rf{qsat1}, \rf{they} and \rf{s6}, using the fact that $\hbP$ and $\hbQ$ commute.  Similarly, the second derivative follows as 
\beq{s8qq}
 \frac{\partial^2 \, }{\partial \phi^2 }{\cal E} ( {\bf n}, \phi) 
= 2\widehat{\bosy{\sigma}}^T\, \big[  \hbP^2({\bf n}) \hbQ({\bf n} , \phi ) \hbS \hbQ^T ({\bf n} , \phi ) + \hbP({\bf n}) \hbQ({\bf n} , \phi ) \hbS \hbQ^T ({\bf n} , \phi ) \hbP^T ({\bf n})
\big]  \,  \widehat{\bosy{\sigma}} 
\, .  \quad
\eeq
%Note that the symmetry and skew symmetry properties of  $\hbS$ and $\hbP$ are not necessary in deriving \rf{s8q} and \rf{s8qq}. 

\subsection{Condition for stationary strain energy}
 
Assume,  with no loss in generality, that the  stationary orientation is at $\phi = 0$. 
If  $\hbS$ is at a stationary point, then the energy should be unchanged regardless of the axis $\bf n$, or 
\beq{s7}
{\cal E} \mbox{ stationary} \quad \Leftrightarrow \quad 
  \left. \frac{\partial \, }{\partial \phi } {\cal E} ( {\bf n}, \phi)\right|_{\phi = 0} = 0, \quad \forall \, |{\bf n} |=1.
\eeq
This becomes, using \rf{s8q} evaluated at  $\phi = 0$,  
\beq{s9}
  \left. \frac{\partial \, }{\partial \phi }{\cal E} ( {\bf n}, \phi)\right|_{\phi = 0} = 
2\widehat{\bosy{\sigma}}^T\,   \hbP({\bf n}) \hbS  \,  \widehat{\bosy{\sigma}}\, .  
\eeq
We now take advantage of the fact that the stress is aligned with the fixed axes.  Thus, eqs.   \rf{a1b} and \rf{s1a} give 
\beq{r1}
\widehat{\bosy{\sigma}}^T\,   \hbP({\bf n})
 = 
 \bigg( 0,\, 0,\, 0,\, \sqrt{2} (\sigma_{\rm III} - \sigma_{\rm II})n_1 ,\,  \sqrt{2} (\sigma_{\rm I} - \sigma_{\rm III})n_2 ,\, 
 \sqrt{2} (\sigma_{\rm II} - \sigma_{\rm I})n_3  \bigg) . 
 \eeq
 Hence, 
\beqa{r2}
  \widehat{\bosy{\sigma}}^T\,   \hbP({\bf n}) \hbS  \,  \widehat{\bosy{\sigma}}\, 
   =   
2  
\left[ \begin{array}{c}
(\sigma_{\rm III} - \sigma_{\rm II})n_1 \\ \\  (\sigma_{\rm I} - \sigma_{\rm III})n_2 \\ \\ (\sigma_{\rm II} - \sigma_{\rm I})n_3
 \end{array} \right]^T\, 
 \left[ \begin{array}{ccc}
 S_{14} &  S_{24} &  S_{34} \\
 && \\ 
 S_{15} &  S_{25} &  S_{35} \\
 && \\
 S_{16} &  S_{26} &  S_{36} 
 \end{array} \right] 
 \left[ \begin{array}{c}
 \sigma_{\rm I} \\  \\ 
 \sigma_{\rm II} \\  \\ 
 \sigma_{\rm III} \end{array} \right] .
   \eeqa
This must vanish for arbitrary direction ${\bf n}$, hence the energy $\cal E$  is stationary if 
\beqa{r3}
   \left[ \begin{array}{ccc}
\sigma_{\rm III} - \sigma_{\rm II} & 0 & 0 \\
 && \\
 0 & \sigma_{\rm I} - \sigma_{\rm III} & 0 \\
 && \\
 0 & 0 & \sigma_{\rm II} - \sigma_{\rm I} \end{array} \right]
 \left[ \begin{array}{ccc}
 S_{14} &  S_{24} &  S_{34} \\
 && \\ 
 S_{15} &  S_{25} &  S_{35} \\
 && \\
 S_{16} &  S_{26} &  S_{36} 
 \end{array} \right] 
 \left[ \begin{array}{c}
 \sigma_{\rm I} \\  \\ 
 \sigma_{\rm II} \\  \\ 
 \sigma_{\rm III} \end{array} \right]  = 
 \left[ \begin{array}{c}
0 \\  \\ 
0 \\  \\ 
0 \end{array} \right] .
   \eeqa

Let us assume for simplicity that the state of stress is triaxial, so that $\sigma_{\rm I},\, \sigma_{\rm II},\, \sigma_{\rm III}$ are distinct.  The left matrix in \rf{r3} can be removed, implying a linear condition in the stress:
\beq{4r}
{\bf E} \, \big(\sigma_{\rm I},\, \sigma_{\rm II},\, \sigma_{\rm III}\big)^T = \big(0,\, 0,\, 0 \big)^T, 
\eeq
where ${\bf E}$ involves moduli (compliances) only
\beqa{r4}
   {\bf E}    = 
 \left[ \begin{array}{ccc}
 S_{14} &  S_{24} &  S_{34} \\
 && \\ 
 S_{15} &  S_{25} &  S_{35} \\
 && \\
 S_{16} &  S_{26} &  S_{36} 
 \end{array} \right] .
   \eeqa
Thus, the energy function ${\cal E}$ is stationary if 
 $\big(\sigma_{\rm I},\, \sigma_{\rm II},\, \sigma_{\rm III}\big)$ is a right null vector of the $3\times 3$ matrix $\bf E$. 
Based on \rf{s3} and \rf{s1}, the condition \rf{4r}  is equivalent to the requirement that the off-diagonal elements of the strain vanish: 
\beq{e12}
\varepsilon_{23}=\varepsilon_{31}=\varepsilon_{12}=0\, ,  
\eeq
or, 
\beqa{r6}
{\cal E} \mbox{ stationary} \quad \Leftrightarrow \quad 
\widehat{\bosy{\varepsilon}}  = \left[ 
\ba{c}
\varepsilon_{1} \\  \varepsilon_{2} \\ \varepsilon_{3} \\ 0 \\ 0 \\ 0
 \ea \right] = \left[ 
\ba{c}
\varepsilon_{\rm I} \\  \varepsilon_{\rm II} \\ \varepsilon_{\rm III} \\ 0 \\ 0 \\ 0
 \ea \right] .
\eeqa 
where $\varepsilon_{\rm I}, \varepsilon_{\rm II}\varepsilon_{\rm III}$ are the  principal strains.  
We have therefore derived the simple but important general result: 

\medskip 
\noindent
{\bf Result 1}: \textcolor{blue}{The energy  
$ \textcolor{blue} {\cal E}$  is stationary iff the stress and strain are coaxial. }
\medskip 

Equation \rf{e12} states that the 3-vector $ (\sigma_{\rm I},\, \sigma_{\rm II},\,\sigma_{\rm III})$ is a right null-vector of ${\bf E}$. 
This  requires as a {\it necessary but not sufficient condition} that 
\beq{v1}
\mbox{det} \, {\bf E} = 0.  
\eeq
Consequences of this condition were explored in detail by Rovati and Taliercio \cite{Rovati03} for particular material symmetries: cubic, transversely isotropic and tetragonal.  A different approach is taken in Section \ref{sec4} below, where the strain energy will be minimized directly.  

While Result 1 is not new but has been derived by several authors 
\cite{Seregin81,Rovati91,Cowin94,Vianello96a,Banichuk96,Rovati03}, 
the present derivation is novel and explicit.  In particular, it allows us to go further and obtain the condition necessary for a minimum or maximum.  This is  explored next. 

\subsection{Condition for an energy  minimum}

The second derivative of $\cal E$ at the stationary point follows from \rf{s8qq} evaluated at $\phi = 0$, 
\beq{cm1}
  \left. \frac{\partial^2 \, }{\partial \phi^2 }{\cal E}(  {\bf n}, \phi)\right|_{\phi = 0} = 
2\widehat{\bosy{\sigma}}^T\, \big[ \hbP^2 \hbS  + \hbP \hbS \hbP^T  \big] \,  \widehat{\bosy{\sigma}}\, .  
\eeq
Each term on the right hand side will be examined in turn.  Using \rf{r1}, it follows that 
\beqa{cm4}
  \widehat{\bosy{\sigma}}^T\, \hbP^2 \hbS  \,  \widehat{\bosy{\sigma}}\, = 2\, 
 %\big( \sigma_{\rm I},\, \sigma_{\rm II},\, \sigma_{\rm III}\big)
   \left[ \begin{array}{c}
 \sigma_{\rm I} \\ \\  \sigma_{\rm II} \\ \\ \sigma_{\rm III}
 \end{array} \right]^T\, 
 \left[ 
\ba{cccccc}
 S_{11} &  S_{12} &  S_{13} & 
    S_{14} &    S_{15} &    S_{16} 
\\ & & & & & \\
 S_{12} &  S_{22} &  S_{23} & 
    S_{24} &    S_{25} &    S_{26} 
\\ & & & & & \\
 S_{13} &  S_{23} &  S_{33} & 
    S_{34} &    S_{35} &    S_{36} 
\ea \right]\, 
\left[ 
\ba{c}
(\sigma_{\rm III} - \sigma_{\rm I})n_2^2 + (\sigma_{\rm II} - \sigma_{\rm I})n_3^2\\
\\
(\sigma_{\rm I} - \sigma_{\rm II})n_3^2 + (\sigma_{\rm III} - \sigma_{\rm II})n_1^2\\
\\
(\sigma_{\rm II} - \sigma_{\rm III})n_1^2 + (\sigma_{\rm I} - \sigma_{\rm III})n_2^2\\
\\
(\sigma_{\rm II} +\sigma_{\rm III} - 2\sigma_{\rm I})n_2n_3 \\
\\
(\sigma_{\rm III} +\sigma_{\rm I} - 2\sigma_{\rm II})n_3n_1 \\
\\
(\sigma_{\rm I} +\sigma_{\rm II} - 2\sigma_{\rm III})n_1n_2 \\
\ea \right],\quad 
\eeqa
and
\beqa{cm2}
  \widehat{\bosy{\sigma}}^T\, \hbP \hbS \hbP^T  \,  \widehat{\bosy{\sigma}}\, =  
   4\,
  \left[ \begin{array}{c}
(\sigma_{\rm III} - \sigma_{\rm II})n_1 \\ \\  (\sigma_{\rm I} - \sigma_{\rm III})n_2 \\ \\ (\sigma_{\rm II} - \sigma_{\rm I})n_3
 \end{array} \right]^T\, 
    \left[ \begin{array}{ccc}
 S_{44} &  S_{45} &  S_{46} \\
 && \\ 
 S_{45} &  S_{55} &  S_{56} \\
 && \\
 S_{46} &  S_{56} &  S_{66} 
 \end{array} \right]\, 
 \left[ \begin{array}{c}
(\sigma_{\rm III} - \sigma_{\rm II})n_1 \\ \\  (\sigma_{\rm I} - \sigma_{\rm III})n_2 \\ \\ (\sigma_{\rm II} - \sigma_{\rm I})n_3
 \end{array} \right]\, . 
\eeqa
Thus, 
\beq{cm6}
   \left. \frac{\partial^2 \, }{\partial \phi^2 }{\cal E} (  {\bf n}, \phi)\right|_{\phi = 0} =   4\, {\bf n}^T {\bf F} {\bf n} , 
\eeq
where the symmetric $3\times 3$ matrix ${\bf F}$ has elements  
\beqa{G}
F_{11} &=& (\sigma_{\rm III} - \sigma_{\rm II})\, \big[ 2 S_{44} (\sigma_{\rm III} - \sigma_{\rm II}) +
	(S_{12}-S_{13})\sigma_{\rm I} + (S_{22}-S_{23})\sigma_{\rm II} + (S_{32}-S_{33})\sigma_{\rm III} \big], 
	\nonumber \\
F_{22} &=& (\sigma_{\rm I} - \sigma_{\rm III})\, \big[ 2 S_{55} (\sigma_{\rm I} - \sigma_{\rm III}) +
	(S_{13}-S_{11})\sigma_{\rm I} + (S_{23}-S_{21})\sigma_{\rm II} + (S_{33}-S_{31})\sigma_{\rm III} \big], 
	\nonumber \\
F_{33} &=& (\sigma_{\rm II} - \sigma_{\rm I})\, \big[ 2 S_{66} (\sigma_{\rm II} - \sigma_{\rm I}) +
	(S_{11}-S_{12})\sigma_{\rm I} + (S_{21}-S_{22})\sigma_{\rm II} + (S_{31}-S_{32})\sigma_{\rm III} \big], 
\nonumber 	 \\
F_{23} &=& 2 S_{56} (\sigma_{\rm I} - \sigma_{\rm III}) (\sigma_{\rm II} - \sigma_{\rm I}) + \frac12 \,
(\sigma_{\rm II} +\sigma_{\rm III}- 2\sigma_{\rm I})\, \big(
	S_{14}\sigma_{\rm I} +S_{24}\sigma_{\rm II} +S_{34}\sigma_{\rm III}\big) ,
\\
F_{31} &=& 2 S_{46} (\sigma_{\rm II} - \sigma_{\rm I}) (\sigma_{\rm III} - \sigma_{\rm II})  + \frac12 \,
(\sigma_{\rm III} +\sigma_{\rm I}- 2\sigma_{\rm II})\, \big(
	S_{15}\sigma_{\rm I} +S_{25}\sigma_{\rm II} +S_{35}\sigma_{\rm III}\big),
\nonumber \\
F_{12} &=& 2 S_{45} (\sigma_{\rm III} - \sigma_{\rm II}) (\sigma_{\rm I} - \sigma_{\rm III}) +\frac12 \,
(\sigma_{\rm I} +\sigma_{\rm II}- 2\sigma_{\rm III})\, \big(
	S_{16}\sigma_{\rm I} +S_{26}\sigma_{\rm II} +S_{36}\sigma_{\rm III}\big)\, . \nonumber
\eeqa
 
The second derivative \rf{cm6} must be positive for all directions ${\bf n}$ at an orientation where $\cal E$ is a local minimum.  Noting that 
\beqa{G4}
F_{23} &=& 2 S_{56} (\sigma_{\rm I} - \sigma_{\rm III}) (\sigma_{\rm II} - \sigma_{\rm I}) + \frac12 \,
(\sigma_{\rm II} +\sigma_{\rm III}- 2\sigma_{\rm I})\, \varepsilon_{23} ,
\nonumber \\
F_{31} &=& 2 S_{46} (\sigma_{\rm II} - \sigma_{\rm I}) (\sigma_{\rm III} - \sigma_{\rm II})  +\frac12 \,
(\sigma_{\rm III} +\sigma_{\rm I}- 2\sigma_{\rm II})\, \varepsilon_{31},
 \\
F_{12} &=& 2 S_{45} (\sigma_{\rm III} - \sigma_{\rm II}) (\sigma_{\rm I} - \sigma_{\rm III}) +\frac12 \,
(\sigma_{\rm I} +\sigma_{\rm II}- 2\sigma_{\rm III})\, \varepsilon_{12} , \nonumber
\eeqa
it follows that at a stationary point the off-diagonal elements of $\bf F$ become, 
\beqa{G5}
F_{23} &=& 2 S_{56} (\sigma_{\rm I} - \sigma_{\rm III}) (\sigma_{\rm II} - \sigma_{\rm I})  ,
\nonumber \\
F_{31} &=& 2 S_{46} (\sigma_{\rm II} - \sigma_{\rm I}) (\sigma_{\rm III} - \sigma_{\rm II})  ,
 \\
F_{12} &=& 2 S_{45} (\sigma_{\rm III} - \sigma_{\rm II}) (\sigma_{\rm I} - \sigma_{\rm III}) . \nonumber
\eeqa
 Equivalently, by pre- and post-multiplication of $\frac 12 \bf F$ by the diagonal matrix
 diag$[(\sigma_{\rm III} - \sigma_{\rm II})^{-1},\, (\sigma_{\rm I} - \sigma_{\rm III})^{-1},\, (\sigma_{\rm II} - \sigma_{\rm I})^{-1}]$, it follows that $\bf G$ must be  positive definite, where 
 \beqa{G6}
G_{11} &=&  S_{44}+ \frac12 (\sigma_{\rm III} - \sigma_{\rm II})^{-1}\, \big[  
	(S_{12}-S_{13})\sigma_{\rm I} + (S_{22}-S_{23})\sigma_{\rm II} + (S_{32}-S_{33})\sigma_{\rm III} \big], 
	\nonumber \\
G_{22} &=&  S_{55} + \frac12 (\sigma_{\rm I} - \sigma_{\rm III})^{-1}\, \big[   
	(S_{13}-S_{11})\sigma_{\rm I} + (S_{23}-S_{21})\sigma_{\rm II} + (S_{33}-S_{31})\sigma_{\rm III} \big], 
	\nonumber \\
G_{33} &=&  S_{66}+ \frac12 (\sigma_{\rm II} - \sigma_{\rm I})^{-1}\, \big[   
	(S_{11}-S_{12})\sigma_{\rm I} + (S_{21}-S_{22})\sigma_{\rm II} + (S_{31}-S_{32})\sigma_{\rm III} \big], 
\nonumber 	 \\
G_{23} &=&  S_{56}  \,  ,
\\
G_{31} &=&  S_{46} \,  ,
\nonumber \\
G_{12} &=&  S_{45} \, . \nonumber
\eeqa
Note that $G_{11} =  S_{44}- \frac12 (\sigma_{\rm III} - \sigma_{\rm II})^{-1}\, (\varepsilon_3 - \varepsilon_2)$, etc. or, using \rf{r6},  
\beqa{G7}
 {\bf G} = \left[ \begin{array}{ccc}
   S_{44}- \frac12 \big( \frac{\varepsilon_{\rm III} - \varepsilon_{\rm II} }{\sigma_{\rm III} - \sigma_{\rm II}}\big)  &   S_{45} &   S_{46} \\
 && \\ 
   S_{45} &  S_{55} -\frac12 \big( \frac{\varepsilon_{\rm I} - \varepsilon_{\rm III} }{\sigma_{\rm I} - \sigma_{\rm III}}\big) &   S_{56}  \\
 && \\ 
   S_{46}&    S_{56} &   S_{66} - \frac12 \big(\frac{\varepsilon_{\rm II} - \varepsilon_{\rm I} }{\sigma_{\rm II} - \sigma_{\rm I}} \big)
  \end{array} \right]\, . 
  \eeqa
In summary, 
\beqa{cmsum}
 \left. \frac{\partial^2 \, }{\partial \phi^2 }{\cal E} (  {\bf n}, \phi)\right|_{\phi = 0}
 =  
   8\,
  \left[ \begin{array}{c}
(\sigma_{\rm III} - \sigma_{\rm II})n_1 \\ \\  (\sigma_{\rm I} - \sigma_{\rm III})n_2 \\ \\ (\sigma_{\rm II} - \sigma_{\rm I})n_3
 \end{array} \right]^T\, 
    {\bf G}\, 
 \left[ \begin{array}{c}
(\sigma_{\rm III} - \sigma_{\rm II})n_1 \\ \\  (\sigma_{\rm I} - \sigma_{\rm III})n_2 \\ \\ (\sigma_{\rm II} - \sigma_{\rm I})n_3
 \end{array} \right]\, . 
\eeqa
This must hold for arbitrary $\bf n$, $|{\bf n}|=1$, and  again assuming that the principal stresses are distinct, it follows that  $\bf G$ must be positive definite.  Combined with Result 1 for the existence of a stationary point, this gives  

\medskip 
\noindent
{\bf Result 2}:  \textcolor{blue}{The energy
${\cal E}$  is a local minimum if the stress and strain are coaxial and the symmetric matrix ${\bf G}$ of \rf{G7} is positive definite.  }  
 
 This can be rewritten (with obvious notation)
\beqa{G7ex}
  \left[ \begin{array}{ccc}
  S_{44} &  S_{45} & S_{46} \\
 && \\ 
S_{45} &  S_{55}  &  S_{56}  \\
 && \\ 
  S_{46}&    S_{56} &   S_{66} 
  \end{array} \right]\, > \, 
  \left[ \begin{array}{ccc}
   \frac12 \big( \frac{\varepsilon_{\rm III} - \varepsilon_{\rm II} }{\sigma_{\rm III} - \sigma_{\rm II}}\big)  &   0 &   0 \\
 && \\ 
   0 &  \frac12 \big( \frac{\varepsilon_{\rm I} - \varepsilon_{\rm III} }{\sigma_{\rm I} - \sigma_{\rm III}}\big) &   0  \\
 && \\ 
   0&    0 &   \frac12 \big(\frac{\varepsilon_{\rm II} - \varepsilon_{\rm I} }{\sigma_{\rm II} - \sigma_{\rm I}} \big)
  \end{array} \right]  \, . 
  \eeqa
 The left matrix is positive definite because of the positive definite properties of the moduli.  The minimum condition therefore requires that this latter matrix be greater than\footnote{The matrix $\bf U$ is greater than the matrix $\bf V$ if 
${\bf r}^T {\bf U} {\bf r} > {\bf r}^T {\bf V} {\bf r}$ for all nonzero  ${\bf r} \in R^3$.}  the right hand diagonal matrix defined by the principal stresses and strains.  The requirement that the full matrix  is positive definite can be relaxed if the stationarity is restricted in  orientation axis $\bf n$.  Thus, only the single scalar quantity ${\bf n}^T{\bf G}{\bf n}$ needs to be considered in the important special case of rotation about a single axis.  This situation is examined in detail in the Appendix.

\section{Optimal orientation of particular material symmetries}\label{sec4}

\subsection{Partition of the energy}
Before considering specific material symmetries, several general results are presented which help focus attention on the anisotropic part of the strain energy.  Separate  contributions to the energy function ${\cal E}$ of \rf{s4} from  isotropic and anisotropic parts of the elastic moduli may be 
distinguished as follows, 
\beqa{fa}
{\cal E}&=& {\cal E}^{\rm (is)} + {\cal E}^{\rm (an)} 
\nonumber \\
 &=&  \sigma_{ij} \sigma_{kl} \, s^{\rm (is)}_{ijkl} +\sigma_{ij} \sigma_{kl} \, s^{\rm (an)}_{ijkl}.  
\eeqa
The moduli are partitioned into isotropic and anisotropic parts 
\beq{wr}
s_{ijkl} =  s^{\rm (is)}_{ijkl} + s^{\rm (an)}_{ijkl},
\qquad
c_{ijkl} =  c^{\rm (is)}_{ijkl} + c^{\rm (an)}_{ijkl},  
\eeq
with the  isotropic moduli defined by 
\beq{wr2}
s^{\rm (is)}_{ijkl} = \frac{1}{3\kappa_s} \, J_{ijkl} + \frac{1}{2\mu_s} \, K_{ijkl},
\qquad
c^{\rm (is)}_{ijkl} = 3\kappa_c \, J_{ijkl} + 2\mu_c \, K_{ijkl}.  
\eeq
Here,   
\beq{wr3}
 J_{ijkl}  = \frac{1}{3}\, \delta_{ij} \delta_{kl}, 
\qquad
K_{ijkl}= I_{ijkl}- J_{ijkl} \, , 
\eeq
and $I_{ijkl}$ are the elements of the fourth order identity. 
The effective ``bulk" and ``shear" moduli $\kappa_s,\, \mu_s$ and $\kappa_c,\, \mu_c$  are obtained as 
\beqa{km}
 && \frac{1}{\kappa_s} =  3 s_{ijkl}J_{ijkl} = 
S_{11}+S_{22}+S_{33}+2S_{12}+2S_{13}+2S_{23}, 
\\
&& \frac{15}{4 \mu_s} =  \frac{3}{2}s_{ijkl}K_{ijkl} = S_{11}+S_{22}+S_{33} - S_{12}-S_{23}-S_{31} +3S_{44}+3S_{55}+3S_{66} ,
\qquad 
\\
&& 9 \kappa_c =  3 c_{ijkl}J_{ijkl} = 
C_{11}+C_{22}+C_{33}+2C_{12}+2C_{13}+2C_{23}, 
\\
&& 15 \mu_c =  \frac{3}{2}c_{ijkl}K_{ijkl} = C_{11}+C_{22}+C_{33} - C_{12}-C_{23}-C_{31} +3C_{44}+3C_{55}+3C_{66} .
\qquad 
\eeqa
Note that in general $\kappa_c \ne \kappa_s$ and $\mu_c\ne \mu_s$.  The anisotropic parts of the moduli in \rf{wr} are defined as the remainder after subtracting the isotropic parts, 
$s^{\rm (an)}_{ijkl} = s_{ijkl} - s^{\rm (is)}_{ijkl}$, etc. 

The energy associated with the isotropic part of the moduli becomes 
\beq{fis}
{\cal E}^{\rm (is)} = \frac{1}{\kappa_s}\overline{\sigma}^2 + \frac{1}{2\mu_s} \sigma_{ij}'\sigma_{ij}',
\eeq
where $\overline{\sigma}$ and ${\bosy{\sigma}}'$ are the hydrostatic and deviatoric stress, respectively, 
\beq{fis2}
\overline{\sigma}=  \frac13 \sigma_{kk}, \qquad \sigma_{ij}' = \sigma_{ij} - \overline{\sigma}\delta_{ij}.
\eeq
These may be written explicitly in terms of the principal stresses, from \rf{s1}, as  
\beqa{sdev}
\overline{\sigma} &=&  \frac13\, \big( \sigma_{\rm I}+\sigma_{\rm II}+\sigma_{\rm III}\big), \\
 \bosy{\sigma}'    &=& \frac13\, 
\left[ \begin{array}{ccc}
2 \sigma_{\rm I}- \sigma_{\rm II}- \sigma_{\rm III} & 0 & 0 \\ && \\
0 & 2 \sigma_{\rm II}- \sigma_{\rm III}- \sigma_{\rm I} &  0 \\ && \\
0 & 0 &  2 \sigma_{\rm III}- \sigma_{\rm I}- \sigma_{\rm II}  
\end{array} \right] 
\equiv \left[ \begin{array}{ccc}
\sigma_{\rm I} ' & 0 & 0 \\ && \\
0 &  \sigma_{\rm II}' &  0 \\ && \\
0 & 0 &   \sigma_{\rm III}'  
\end{array} \right] .\qquad 
\eeqa 
The energy associated with  the anisotropic part of the moduli is
\beq{fan}
{\cal E}^{\rm (an)} = \overline{\sigma}^2 s^{\rm (an)}_{jjkk} + 2\overline{\sigma} \sigma_{ij}' \, s^{\rm (an)}_{ijkk} + 
\sigma_{ij}' \sigma_{kl}' \, s^{\rm (an)}_{ijkl} .
\eeq
By definition, the scalar quantity $s^{\rm (an)}_{jjkk}$ is zero, and accordingly, the anisotropic energy simplifies to 
\beq{fq}
{\cal E}^{\rm (an)} =2\overline{\sigma} \sigma_{ij}' \, s^{\rm (an)}_{ijkk} + 
\sigma_{ij}' \sigma_{kl}' \, s^{\rm (an)}_{ijkl} .
\eeq

It may be shown that the $3\times 3$ matrices $\bf E$ and $\bf G$ of \rf{r4} and \rf{G7} vanish  for isotropic materials.  In general, they depend upon the anisotropic part of the material moduli.

\subsection{Materials with cubic symmetry}  

In the fixed coordinate system of the principal stress axes, the  elastic compliance for  a material with cubic symmetry is 
\beqa{cub1}
\cS^{(0)}  = \left[ 
\ba{cccccc}
S_{11}^{(0)} & S_{12}^{(0)} & S_{12}^{(0)} & 
 0&0&0 
\\ & & & & & \\
 & S_{11}^{(0)} & S_{12}^{(0)} & 
 0&0&0
\\ & & & & & \\
 & & S_{11}^{(0)} & 
 0&0&0
\\ & & & & & \\
 &   &   
 & S_{44}^{(0)} & 0 & 0
\\ & & & & & \\
   S &Y  &M & & S_{44}^{(0)}&  0 
\\ & & & & & \\
&&&&& S_{44}^{(0)}
\ea \right] . 
\eeqa
There are three independent  moduli, $\kappa$, $\mu_1$ and $\mu_2$, where
\beq{cub2}
\frac{1}{3\kappa} =   S_{11}^{(0)}+2  S_{12}^{(0)},\quad 
\frac{1}{2\mu_1} = 2S_{44}^{(0)},\quad
\frac{1}{2\mu_2} =   S_{11}^{(0)}-  S_{12}^{(0)}\, . 
\eeq 
The associated fourth order tensors  can be expressed succinctly using the irreducible tensor  notation of Walpole \cite{c1}, as
\beq{wr5}
c^{(0)}_{ijkl} = 3\kappa \, J_{ijkl} + 2\mu_1 \, L^{(0)}_{ijkl}+ 2\mu_2 \, M^{(0)}_{ijkl}, 
\qquad
s^{(0)}_{ijkl} = \frac{1}{3\kappa} \, J_{ijkl} + \frac{1}{2\mu_1} \, L^{(0)}_{ijkl}+ \frac{1}{2\mu_2} \, M^{(0)}_{ijkl}. 
\eeq
Here $L^{(0)}_{ijkl}= I_{ijkl}- D^{(0)}_{ijkl}$,  $M^{(0)}_{ijkl}= D^{(0)}_{ijkl}- J_{ijkl}$, and 
\beq{Ddef}
D^{(0)}_{ijkl} = \delta_{i1} \delta_{j1}\delta_{k1} \delta_{l1}
+\delta_{i2} \delta_{j2}\delta_{k2} \delta_{l2} +\delta_{i3} \delta_{j3}\delta_{k3} \delta_{l3}\, . 
\eeq
This format makes it relatively straightforward to determine the effective isotropic moduli, 
\beq{wr6}
\kappa_c = \kappa_s = \kappa, \qquad 5\mu_c = 3\mu_1+2\mu_2, \qquad
\frac{5}{\mu_s} = \frac{3}{\mu_1} + \frac{2}{\mu_2}\, . 
 \eeq
Thus, 
\beq{w8}
s^{(an,0)}_{ijkl} = \frac{1}{10}\bigg(\frac{1}{\mu_1}- \frac{1}{\mu_2}\bigg)\, \big(
5J_{ijkl} + 2K_{ijkl} - 5D^{(0)}_{ijkl} \big). 
\eeq

The anisotropic  part of the energy, \rf{fq}, depends only upon the deviatoric part of the stress,  
\beq{fq2}
{\cal E}^{\rm (an)} =\sigma_{ij}' \sigma_{kl}' \, s^{\rm (an)}_{ijkl} .
\eeq
The reason is that the second order tensor $s^{(an,0)}_{ijkk}$ is identically zero for cubic symmetry, and hence  remains zero in the rotated material axes: $s^{\rm (an)}_{ijkk} = 0$.  The first term in \rf{fq} therefore vanishes, leaving the simpler expression \rf{fq2}.  The isotropic tensors $J_{ijkl}$ and $K_{ijkl}$ are  unchanged under rotation, and consequently, from \rf{fis}, \rf{w8} and \rf{fq2}, 
\beq{fq3}
{\cal E} = \frac{1}{\kappa}\overline{\sigma}^2 + \frac{1}{2\mu_1} \sigma_{ij}'\sigma_{ij}' + {\cal E}^{\rm (ex)},
\qquad
{\cal E}^{\rm (ex)} = \frac{1}{2}\bigg(\frac{1}{\mu_2}- \frac{1}{\mu_1}\bigg)\,
\sigma_{ij}' \sigma_{kl}' \, D_{ijkl} ,  
\eeq
where $D_{ijkl}$ is the rotated version of $D^{(0)}_{ijkl}$.  In order to avoid ambiguity, let $\sigma_{(rot)kl}$ denote the stress in the {\it rotated} coordinate system, then it follows that
\beq{fq4}
\sigma_{ij}' \sigma_{kl}' \, D_{ijkl} = \sigma_{(rot)11}^{'2} + \sigma_{(rot)22}^{'2} +\sigma_{(rot)33}^{'2}  \ . 
\eeq
The scalar second invariant of the deviatoric stress is 
\beqa{in}
\sigma_{ij}'\sigma_{ij}' &=& {\sigma_{\rm I}'}^2 + {\sigma_{\rm II}'}^2 + {\sigma_{\rm III}'}^2
\nonumber \\
&=& \sigma_{(rot)11}^{'2} + \sigma_{(rot)22}^{'2} +\sigma_{(rot)33}^{'2}
+ 2\sigma_{(rot)23}^{'2} + 2\sigma_{(rot)31}^{'2} +2\sigma_{(rot)12}^{'2} .  
\eeqa
Therefore, the function ${\cal E}^{\rm (ex)}$ of \rf{fq3} is stationary when either the right member of \rf{fq4} or 
\beq{fq6}
\Gamma = \sigma_{(rot)23}^{'2} + \sigma_{(rot)31}^{'2} +\sigma_{(rot)12}^{'2},
\eeq
are stationary.  Furthermore, 
\beq{fq7}
{\cal E}^{\rm (ex)} ={\cal E}^{\rm (ex,0)}  +  \bigg(\frac{1}{\mu_1}- \frac{1}{\mu_2}\bigg)\,
\big[ \sigma_{(rot)23}^{'2} + \sigma_{(rot)31}^{'2} +\sigma_{(rot)12}^{'2}\big] \, , 
\eeq
where ${\cal E}^{\rm (ex,0)}$ is the unrotated or fixed value, which follows from \rf{sdev} as
\beq{q9}
{\cal E}^{\rm (ex,0)} = \frac{1}{2}\bigg(\frac{1}{\mu_2}- \frac{1}{\mu_1}\bigg)\,  
\big( {\sigma_{\rm I}'}^2 + {\sigma_{\rm II}'}^2 + {\sigma_{\rm III}'}^2 \big)\, . 
\eeq
 Hence, 
\beq{wq3}
 \mu_1 > \mu_2 \quad \Rightarrow \quad {\cal E}^{\rm (ex)}\le  {\cal E}^{\rm (ex,0)}, 
 \eeq
with equality when the material and stress axes are aligned.  Thus, \textcolor{blue}{a local minimum that is not aligned with the stress axes occurs iff $\mu_1 > \mu_2$ and occurs when $\Gamma$ of \rf{fq6} achieves a local maximum}.   It will be shown below that the maximum value is $\frac12 \sigma_{ij}'\sigma_{ij}'$, or equivalently, that ${\cal E}^{\rm (ex)}$ is zero at the stationary point.

As  the material axes are rotated to transform $s^{(an,0)}_{ijkl} \rightarrow s^{\rm (an)}_{ijkl}$, the only part that contributes to the anisotropic strain energy  is $D_{ijkl}^{(0)} \rightarrow D_{ijkl}$.  
Conditions for obtaining the stationary value of strain energy are next derived by focusing on the dependence upon $D_{ijkl}$.  The $6\times 6$   matrix associated with the unrotated tensor $D^{(0)}_{ijkl}$ is
\beqa{D1}
\widehat{\bf D}^{(0)}    = 
\left[ \begin{array}{ccc}
I_{3\times 3} & &0_{3\times 3}  \\ && \\ 
0_{3\times 3} && 0_{3\times 3} 
\end{array} \right] .
\eeqa 
It is convenient to split $ \hbQ $ as follows into $3\times 3$ matrices, 
\beqa{D2}
 \hbQ   = 
\left[ \begin{array}{ccc}
\hbQ_1 & &\hbQ_2  \\ && \\ 
\hbQ_3 && \hbQ_4 
\end{array} \right] ,   
\eeqa 
so that the rotated tensor    
$\widehat{{\bf D}}  = \hbQ \widehat{{\bf D}}^{(0)} \hbQ^T$,  follows from \rf{D1} and \rf{D2} as 
\beqa{D2a}
\widehat{{\bf D}}  = 
\left[ \begin{array}{ccc}
\hbQ_1 \hbQ_1^T & &\hbQ_1 \hbQ_3^T  \\ && \\ 
\hbQ_3 \hbQ_1^T && \hbQ_3 \hbQ_3^T 
\end{array} \right] .  
\eeqa 
The term associated with the rotated energy  follows from \rf{s1a} and \rf{D2a} as  
\beq{dq2}
\sigma_{ij}' \sigma_{kl}' \, D_{ijkl} =   (\sigma_{\rm I}',\, \sigma_{\rm II}',\,\sigma_{\rm III}') \, \hbQ_1\hbQ_1^T \, (\sigma_{\rm I}',\, \sigma_{\rm II}',\,\sigma_{\rm III}')^T\, .   
\eeq
  Thus,  any  stress that is a null vector of $\hbQ_1^T$
also  yields the minimum or maximum value for ${\cal E}^{\rm (ex)}$ of \rf{fq3}, i.e. zero.  
This suggests $\hbQ_1$ as the focus of attention, and implies    
 the important result: \textcolor{blue}{Every stress state which is a null vector of $\hbQ_1^T$ corresponds to a global minimum (maximum) of ${\cal E}$ if $\mu_1 >\mu_2$ ($\mu_2 >\mu_1$)}.  We therefore search for null vectors of $\hbQ_1^T$. 

Before deriving two alternative methods to find null vectors of $\hbQ_1^T$ in the next two subsections, note that the quantity \rf{fq4} vanishes at a stationary orientation, and hence $\sigma_{(rot)11} = \sigma_{(rot)22}  = \sigma_{(rot)22}$.  Thus, the  stress in each of the three rotated axial directions is equal, a result previously obtained by Rovati and Taliercio \cite{Rovati91,Rovati03}.   Furthermore, at the stationary point it may be easily shown that the following identities hold:
\beq{idh}
{\cal D}\bosy{\sigma} = \overline{\sigma}\, {\bf I} , \quad
{\cal L}\bosy{\sigma} = \bosy{\sigma}', \quad 
{\cal M}\bosy{\sigma} = 0\, ,
\eeq
where ${\cal D}, {\cal L}, {\cal M}$ are the (rotated) tensors with components $D_{ijkl}, L_{ijkl}, M_{ijkl}$, respectively.  Hence, the strain at optimal orientation is simply
\beq{idh2}
\bosy{\varepsilon} = \frac{1 }{3\kappa} \, \overline{\sigma} \, {\bf I}   \quad
+ \frac{ 1 }{2\mu_1}\, \bosy{\sigma}'\, \qquad (\mbox{optimal orientation only}).
\eeq
This is clearly coaxial with the stress, which follows from the commutation property of coaxial second order tensors 
($\bosy{\sigma} \bosy{\varepsilon} - \bosy{\varepsilon}  \bosy{\sigma}=0$ in this case). 

It is also worth remarking that we do not seek null vectors of  the matrix  $\bf E$, although this approach is feasible and has been used to  advantage by Rovati and Taliercio \cite{Rovati03}. Some comments on $\bf E$ are in order though.  The $3\times 3$ matrix follows from  eqs. \rf{D1} through \rf{D2a} as 
\beq{D3}
{\bf E} = \hbQ_3\hbQ_1^T\, ,  
\eeq
and the condition \rf{v1}  is satisfied if either  $\mbox{det} \, \hbQ_1$ or $\mbox{det} \, \hbQ_3$ vanish.  These can be made more explicit in terms of the elements of the rotation matrix.  
Let 
\beqa{qq}
{\bf Q} = 
\left[ \begin{array}{ccc}
q_{11} & q_{12} & q_{13} \\
q_{21} & q_{22} & q_{23} \\
q_{31} & q_{32} & q_{33} 
\end{array} \right] ,  
\eeqa
then using the   the explicit representation of the $6\times 6$ rotation matrix from Auld  \cite{AuldI} or otherwise, the condition  \rf{v1} implies 
\beqa{D5}
\mbox{det} \, \hbQ_1 = \left| \begin{array}{cccccc}
q_{11}^2 & q_{12}^2 & q_{13}^2    \\
&& \\
q_{21}^2 & q_{22}^2 & q_{23}^2    \\
&& \\
q_{31}^2 & q_{32}^2 & q_{33}^2  
\end{array} \right| = 0,
\quad 
\mbox{or}\quad    \mbox{det} \, \hbQ_3 = (2)^{3/2}\, 
\left| \begin{array}{cccccc}
 q_{21}q_{31} & q_{22}q_{32}& q_{23}q_{33} \\
&& \\
 q_{31}q_{11} & q_{32}q_{12}& q_{33}q_{13} \\
&& \\
 q_{11}q_{21} & q_{12}q_{22}& q_{13}q_{23}
\end{array} \right| = 0.\quad
\eeqa
Using the fact that the column vectors of a transformation matrix form an orthonormal triad, it follows that 
\beq{D7a}
 \hbQ_1^T\, (1,\, 1,\, 1)^T =  (1,\, 1,\, 1)^T, \quad  \hbQ_3\,(1,\, 1,\, 1)^T =  (0,\, 0,\, 0)^T.   
\eeq
That is, $(\sigma_{\rm I},\, \sigma_{\rm II},\,\sigma_{\rm III}) = \lambda\, (1,\, 1,\, 1)$ is a   null vector of $\bf E$ for any $\lambda$.  Hence, $\bf E$ is not of full rank, implying that det$\, {\bf E}$ is always zero.  However, this   is not of interest as the null space corresponds to hydrostatic stress, for which the energy function is independent of material orientation.  The implications of \rf{D5} are not considered further, and we return to the simpler task of finding null vectors of $\hbQ_1^T$ alone.

Two  methods for achieving the minimum energy ${\cal E}^{\rm (ex)} = 0$ are  described, both using explicit forms of the rotation.  The first involves a single rotation about an arbitrary axis, and the second is in terms of standard Euler angles.   

\subsubsection{Minimum energy state with a single rotation}

The range of transformations which correspond to energy minima can be obtained using Euler's Theorem \cite{Baruh} which states that any transformation matrix ${\bf Q}$ can be represented in the form \rf{qeq} for some axis ${\bf n}$, $|{\bf n}|=1$,  and angle $\phi$.  Thus, 
\beqa{qeq1}
{\bf Q}({\bf n},\, \phi) = 
\left[ \begin{array}{ccc}
 		1-2s^2(n_2^2+n_3^2) & 2s(sn_1n_2-cn_3) & 2s(sn_1n_3+cn_2)
  \\ && \\
 	 2s(sn_1n_2+cn_3)	& ~ 1-2s^2(n_3^2+n_1^2) ~&  2s(sn_2n_3 -cn_1)
 \\ && \\
 	 2s(sn_1n_3 -cn_2)   &  2s(sn_2n_3+cn_1) & 1-2s^2(n_1^2+n_2^2)  
\end{array} \right] ,\quad
\eeqa
where $c=\cos \frac{\phi}{2}$, $s=\sin \frac{\phi}{2}$.
and the elements of the associated $\hbQ_1$ are determined by squaring each element in \rf{qeq1}. It may be shown that 
\beq{qeq2}
\mbox{det} \, \hbQ_1\ ({\bf n},\, \phi) = \cos 2\phi + 2(2+ 3 \cos\phi) (1-\cos\phi)^2 \,  (
n_1^2n_2^2 + n_2^2n_3^2 + n_3^2n_1^2 )+ 6(1-\cos\phi)^3 \,  
n_1^2n_2^2 n_3^2 \, .  \quad   
\eeq
Note that $n_1^2n_2^2 + n_2^2n_3^2 + n_3^2n_1^2 \le 1/3$ and $n_1^2n_2^2 n_3^2 \le 1/27$ with equality when $n_1^2= n_2^2 =n_3^2 = 1/3$.  

For a given $n_3^2$ and angle $\phi$, 
\beq{n1n2}
n_1^2, n_2^2 =  \frac{1}{2} (1-n_3^2) \pm \big[ \frac{1}{4} (1-n_3^2)^2- g\big]^{1/2}, 
\eeq
where 
\beq{geq}
g(n_3^2, \phi) = 
\frac{-\cos 2\phi - 2(2+ 3 \cos\phi) (1-\cos\phi)^2 \,  n_3^2(1-n_3^2 )}{ 
2 (1-\cos\phi)^2 [ (2+ 3 n_3^2 +3\cos\phi(1 - n_3^2)]}\, .    
\eeq

The null vector of $\hbQ_1$ is such that 
\beqa{nv1}
\sigma_{\rm I} ' Q_{11}^2 + \sigma_{\rm II} ' Q_{12}^2 + \sigma_{\rm III} ' Q_{13}^2 &=&0, 
\nonumber \\
\sigma_{\rm I} ' Q_{21}^2 + \sigma_{\rm II} ' Q_{22}^2 + \sigma_{\rm III} ' Q_{23}^2 &=&0,
\\
\sigma_{\rm I} ' Q_{31}^2 + \sigma_{\rm II} ' Q_{32}^2 + \sigma_{\rm III} ' Q_{33}^2 &=&0.
\nonumber
\eeqa
Using the fact that this is a deviatoric stress, we replace  $ \sigma_{\rm III} ' = -  \sigma_{\rm I} '- \sigma_{\rm II} '$ in the final equation of \rf{nv1}, to get
\beq{nv2}
\sigma_{\rm I} ' \bigg(Q_{31}^2 - Q_{33}^2\bigg) + \sigma_{\rm II} ' \bigg(Q_{32}^2 - Q_{33}^2\bigg) = 0.
\eeq
Hence, 
\beq{nv3}
\sigma_{\rm I} '  = a_0 \bigg(Q_{32}^2 - Q_{33}^2\bigg) ,\quad  \sigma_{\rm II} ' = a_0\bigg(Q_{33}^2 - Q_{31}^2\bigg) ,
\eeq
for arbitrary  $a_0 \ne 0$. Once again using the fact that $ \sigma_{\rm III} ' = -  \sigma_{\rm I} '- \sigma_{\rm II} '$ gives
\beq{nv3a}
\sigma_{\rm III} '  = a_0 \bigg(Q_{31}^2 - Q_{32}^2\bigg) . 
\eeq
In the same way, using the other equations in \rf{nv1}, three alternative expressions for the null vector are found:
\beqa{3alt}
\big(\sigma_{\rm I} ',\, \sigma_{\rm II} ',\,  \sigma_{\rm III} '\big)  
&=& a_1 \bigg(Q_{32}^2 - Q_{32}^2,\,  Q_{33}^2 - Q_{31}^2,\, Q_{31}^2 - Q_{32}^2 \bigg) \\
&=& a_2 \bigg(Q_{12}^2 - Q_{12}^2,\,  Q_{13}^2 - Q_{11}^2,\, Q_{11}^2 - Q_{12}^2 \bigg) \\
&=& a_3 \bigg(Q_{22}^2 - Q_{22}^2,\,  Q_{23}^2 - Q_{21}^2,\, Q_{21}^2 - Q_{22}^2 \bigg) ,  
\eeqa
for some constants $a_1$, $a_2$, $a_3$.  Thus, from the first expression, with $a_1=1$, 
\beqa{s3a}
\sigma_{\rm I} '  &=&  \big[   (1+n_1^2)(n_2^2-n_3^2)(1-\cos \phi) +n_3^2-n_2^2 - 4 n_1n_2n_3\sin\phi\big](1-\cos \phi) , \\
\sigma_{\rm II} '  &=&   \big\{   [(1-n_1^2)(n_1^2-n_3^2) -2n_2^2](1-\cos \phi) +n_2^2-n_1^2 +1 +2 n_1n_2n_3\sin\phi\big\}(1-\cos \phi) - 1, \\
\sigma_{\rm III} '  &=&   \big\{   [(1-n_1^2)(n_2^2-n_1^2) +2n_3^2](1-\cos \phi) +n_1^2-n_3^2 -1 +2 n_1n_2n_3\sin\phi \big\}(1-\cos \phi) + 1, \qquad \label{s3c}
\eeqa
These equations provide us with a two parameter set of stress states, described by $0< n_3^2< 1 $ and $\phi$.  The two are independent insofar as  $n_1^2$ and $n_2^2$ of \rf{n1n2} lie in $(0,1)$.  This in turn requires that $g$ of \rf{geq} satisfies 
\beq{n1n2q}
0 \le g(n_3^2, \phi) \le  \frac{1}{4} (1-n_3^2)^2, 
\eeq
which defines the range of $0< n_3^2< 1 $ and $\phi$. 

\subsubsection{Minimum energy using Euler angles}

The standard three Euler angles  $(\theta_1 ,\, \theta_2 ,\, \theta_3)$ are used to transform from 
$	\{ {\bf e}_1, {\bf e}_2, {\bf e}_3\} 
\rightarrow 
	\{ {\bf e}_1', {\bf e}_2', {\bf e}_3' = {\bf e}_3\}
\rightarrow 
	\{ {\bf e}_1''={\bf e}_1', {\bf e}_2'', {\bf e}_3'' \}
\rightarrow 
	\{ {\bf e}_1''', {\bf e}_2''', {\bf e}_3'''={\bf e}_3'' \}
$.  That is,  first rotate about the ${\bf e}_3$ axis by $\theta_1$, then about the ${\bf e}_1'$ axis by $\theta_2$, and finally about the ${\bf e}_3''$ axis by $\theta_3$.  The transformation matrix is 
\beqa{eul}
&&{\bf Q} (\theta_1,\theta_2,\theta_3) \nonumber \\ && \nonumber \\ && =  
\left[ \begin{array}{ccc}
 		\cos \theta_1 \cos \theta_3 - \sin \theta_1 \cos \theta_2 \sin \theta_3 
 & 	\sin \theta_1\cos \theta_3 + \cos \theta_1 \cos \theta_2  \sin \theta_3
 & 	\sin \theta_2 \sin \theta_3
  \\ && \\
 		-\cos \theta_1\sin \theta_3 - \sin \theta_1 \cos \theta_2 \cos \theta_3 
 & 	-\sin \theta_1\sin \theta_3 + \cos \theta_1 \cos \theta_2 \cos \theta_3 
 & 	\sin \theta_2 \cos \theta_3
 \\ && \\
 	\sin \theta_1 \sin \theta_2 
 & 	-\cos \theta_1 \sin \theta_2  & \cos \theta_2
\end{array} \right] , \qquad
\eeqa
and it follows from this and $\rf{D5}_1$  that 
\beq{detex}
\mbox{det} \, \hbQ_1\ (\theta_1,\theta_2,\theta_3) = \cos 2\theta_1\, \cos 2\theta_2\, \cos 2\theta_3
-\frac14\, \sin 2\theta_1\,  \sin 2\theta_3\, \cos \theta_2\, (3\cos 2 \theta_2 + 1)\, .    \quad
\eeq
The condition that this vanish  is equivalent to eq. (90) of Rovati and Taliercio \cite{Rovati03}, although their result is obtained in a different manner.

Consider, for instance,  $\theta_3=0$, for which 
\beq{D8}
\mbox{det} \, \hbQ_1\ (\theta_1,\theta_2,0) = \cos 2\theta_1\, \cos 2\theta_2,   
\eeq
and hence there are null spaces associated with $\hbQ_1(\theta_1, \pi/4,0)$ and $\hbQ_1(\pi/4,\theta_2,0)$.  The  null spaces are lines in the stress space, 
which follow from the simplified form of $\hbQ_1^T$, 
\beqa{td}
\hbQ_1^T (\theta_1,\theta_2,0) = 
\left[ \begin{array}{ccc}
 \cos^2 \theta_1 ~&~  \sin^2 \theta_1 \cos^2 \theta_2  ~&~  \sin^2 \theta_1  \sin^2 \theta_2
  \\ && \\
 \sin^2 \theta_1     & \cos^2 \theta_1\cos^2 \theta_2  
 & \cos^2 \theta_1 \sin^2 \theta_2
 \\ && \\
0 &  \sin^2 \theta_2  & \cos^2 \theta_2
\end{array} \right] . \quad
\eeqa
The possible states of deviatoric stress are:  
$(\sigma_{\rm I}',\, \sigma_{\rm II}',\,\sigma_{\rm III}') = \lambda \, (0, \, -1,\, 1)$ if $\theta_2 = \pi/4,\, \theta_3 = 0$, and 
$(\sigma_{\rm I}',\, \sigma_{\rm II}',\,\sigma_{\rm III}') = \lambda \, (\cos 2\theta_2, \, -\cos^2\theta_2,\, 
\sin^2\theta_2)$, if $\theta_1 = \pi/4,\, \theta_3 = 0$. The first family of stresses correspond to a 2D elasticity problem (see Appendix):  $\sigma_{\rm I}'=0,\, \sigma_{\rm II}'+ \sigma_{\rm III}'=0$, and it is also a null vector of $\hbQ_1 (0,  \pi/4,\theta_3)$.  The second is also a null vector of $\hbQ_1 (0, \theta_2 , \pi/4)$.  Similarly, $\lambda \, ( 1,\, -1,\, 0)$ is a null vector of $\hbQ_1( \pi/4,\pi/2, \theta_3)$ and 
$\lambda \, ( -\cos^2\theta_1,\, \sin^2\theta_1 , \cos 2\theta_1 )$ is a null vector of $\hbQ_1( \theta_1, \pi/2,\pi/4)$. 

Conversely, an orientation which provides a minimum in energy can be found for a given state of stress.  Assume with no loss in generality that 
$\sigma_{\rm II}' < 0 < \sigma_{\rm III}'$.  Define the angle  $\theta_2$ by 
\beq{lp}
\tan^2 \theta_2 = -\frac{\sigma_{\rm III}'}{\sigma_{\rm II}'},
\eeq
then the deviatoric stress may be expressed
\beq{exp}
(\sigma_{\rm I}',\, \sigma_{\rm II}',\,\sigma_{\rm III}') = \big( \sigma_{\rm III}' - \sigma_{\rm II}'\big)\, (\cos 2\theta_2, \, -\cos^2\theta_2,\, 
\sin^2\theta_2). 
\eeq
It follows from the above example  that this deviatoric stress is a null vector of $\hbQ_1(\pi/4,\theta_2,0)$.

It is instructive to  examine further the example of eq. \rf{exp}.  The rotated material axes,  denoted  $\{ {\bf e}_1', {\bf e}_2',{\bf e}_3'\}$, are given by the columns of ${\bf  Q}(\pi/4,\theta_2,0)$: 
\beqa{inst1}
{\bf e}_1' = \frac{1}{\sqrt{2}}
\left[ \begin{array}{r}
 1 \\ \\
 -c  \\ \\
 s   
\end{array} \right], 
\quad
{\bf e}_2' = \frac{1}{\sqrt{2}}
\left[ \begin{array}{r}
 1  \\ \\
 c  \\ \\
 -s   
\end{array} \right], 
\quad
{\bf e}_3' = 
\left[ \begin{array}{c}
0 \\ \\
 s \\ \\
 c  
\end{array} \right], 
\quad
\eeqa
where $s=\sin \theta_2$, $c=\cos \theta_2$, or from \rf{lp},
\beq{inst2}
s = \sqrt{ \frac{\sigma_{\rm III}'}{\sigma_{\rm III}' - \sigma_{\rm II}'}}, 
\quad
c = \sqrt{ \frac{-\sigma_{\rm II}'}{\sigma_{\rm III}' - \sigma_{\rm II}'}}. 
\eeq
The rotated tensor $D_{ijkl}$ is
\beq{inst3}
{\cal D} = {\bf e}_1'\otimes{\bf e}_1'\otimes{\bf e}_1'\otimes{\bf e}_1'
+ {\bf e}_2'\otimes{\bf e}_2'\otimes{\bf e}_2'\otimes{\bf e}_2'
+ {\bf e}_3'\otimes{\bf e}_3'\otimes{\bf e}_3'\otimes{\bf e}_3'\, , 
\eeq
and hence
\beq{inst4}
{\cal D} \bosy{\sigma}' = \sum \limits_{k=1}^3 {\bf e}_k'\otimes{\bf e}_k'\, 
 ({\bf e}_k'\cdot \bosy{\sigma}' {\bf e}_k' )\, , 
\eeq
It may be seen by direct calculation that the three scalars ${\bf e}_k'\cdot \bosy{\sigma}' {\bf e}_k'$ (no sum) are identically zero by virtue of \rf{inst1} and \rf{inst2}.  This demonstrates  explicitly that 
\beq{inst5}
{\cal D} \bosy{\sigma}' = 0 , 
\eeq
at the optimal orientation.  The identities \rf{idh} follow accordingly. 

\subsubsection{Summary for cubic symmetry}

 The extreme values of the energy for cubic materials are ${\cal E}_1$ and ${\cal E}_2$, where 
\beq{ex12}
{\cal E}_j = \frac{1}{\kappa}\overline{\sigma}^2 + \frac{1}{2\mu_j}
\big( {\sigma_{\rm I}'}^2 + {\sigma_{\rm II}'}^2 + {\sigma_{\rm III}'}^2 \big), \qquad j=1,\, 2. 
\eeq  
The 
fixed axes are always one of the stationary orientations, since $\bf E$ of \rf{r4} vanishes.  The stationary value for the unrotated axes is ${\cal E}_2$, which is the global minimum (maximum) if $\mu_2 > \mu_1$ ($\mu_2 < \mu_1$). The stationary value ${\cal E}_1$ occurs at some rotated axes, 
the existence of which is not in doubt for a material of cubic symmetry (or any material symmetry for that matter). The important point to note is that it is possible to explicitly determine  such orientations. Thus, we have shown  
by direct construction the material orientation that achieves the stationary energy value ${\cal E}_1$ for any state of stress.  This is a global minimum (maximum) if $\mu_2 < \mu_1$ ($\mu_2 > \mu_1$).   

It is interesting that the expressions for the extreme values in  \rf{ex12} have the form of the energy for an isotropic solid, but with different shear moduli.  This is evident by writing ${\cal E}_1$ and ${\cal E}_2$ in terms of the invariants of the stress tensor: 
\beq{ex122}
{\cal E}_j = \frac{1}{9\kappa}\big({\rm tr}\, \bosy{\sigma}\big)^2
+ \frac{1}{2\mu_j}\big[ {\rm tr}\, \bosy{\sigma}^2  - \frac13 \big({\rm tr}\, \bosy{\sigma}\big)^2 \big], \qquad j=1,\, 2. 
\eeq

\subsubsection{Example materials}
Noting that $\mu_1 = c_{44}$ and $\mu_2 = (c_{11}-c_{12})/2$  allows us to determine the sign of $(\mu_2 - \mu_1)$. 
Table A.1 of Musgrave \cite{Musgrave} provides data for $c^* = 2(\mu_2 - \mu_1)$ for a multitude of materials.  These show $c^*$ to be negative for most elemental and engineering materials with cubic symmetry and different lattice structures: aluminum, nickel, copper, silver, gold (all f.c.c structure), iron (b.c.c), brass (f.c.c. and b.c.c), diamond, silicon, germanium (all diamond structure), and GaSb, InSb (both zinc-blende).  Hence, for all of these cubic materials there exist optimal orientations of the axes that achieve the lowest energy state possible.  Materials with positive $c^*$ include crystalline compounds of potassium, sodium and silver with rock-salt structure: KF, KCL, KBr, KI, NaF, NaCl, NaBr, NaI, AgCl, AgBr; plus  caesium compounds with structure related to b.c.c.  For these, the orientation associated with \rf{exp}, for instance, gives maximum strain energy.  The minimum energy is achieved by no rotation. 

\subsection{Transverse isotropy} 
Materials with hexagonal symmetry, or equivalently, transverse isotropy (TI), are characterized by five moduli.  In the coordinate system of the principal axes, the elements of the compliance are  
\beqa{TI1}
\cS^{(0)}  = \left[ 
\ba{cccccc}
S_{11}^{(0)} & S_{12}^{(0)} & S_{13}^{(0)} & 
 0&0&0 
\\ & & & & & \\
 & S_{11}^{(0)} & S_{13}^{(0)} & 
 0&0&0
\\ & & & & & \\
 & & S_{33}^{(0)} & 
 0&0&0
\\ & & & & & \\
 &   &   
 & S_{44}^{(0)} & 0 & 0
\\ & & & & & \\
   S &Y  &M & & S_{44}^{(0)}&  0 
\\ & & & & & \\
&&&&& S_{66}^{(0)}
\ea \right] ,
\eeqa
with $S_{66}^{(0)} = \frac12 ( S_{11}^{(0)} - S_{12}^{(0)} )$.  The TI material is characterized by an axis of symmetry, defined by the unit vector ${\bf n}$, which is here chosen as the ${\bf e}_3$ axis.  In general, the strain energy depends only upon the orientation of ${\bf n}$ with respect to the stress axes, and  the problem is  formulated as one of selecting ${\bf n}$ to minimize $\cal E$. 

First note that two of the five moduli can be ascribed to the isotropic part of the elasticity; or conversely, an isotropic part   may be subtracted from the compliance  tensor $s_{ijkl}$ according to \rf{wr}, \rf{wr2} and \rf{km}, where 
 \beq{km3}
 \frac{1}{\kappa_s} =  
2S_{11}^{(0)}+S_{33}^{(0)}+2S_{12}^{(0)}+4S_{13}^{(0)}, \qquad
 \frac{15}{4 \mu_s} =   2S_{11}^{(0)}+S_{33}^{(0)} - S_{12}^{(0)} - 2S_{13}^{(0)}+6S_{44}^{(0)}+3S_{66}^{(0)} ,
 \eeq
  leaving a tensor $s_{ijkl}^{\rm (an)}$ with three constants.  The anisotropic compliance depends upon the orientation of the axis of symmetry as follows  
\beqa{t1}
s_{ijkl}^{\rm (an)} &=& a\, n_in_jn_kn_l + b\, (\delta_{ij}n_kn_l+ \delta_{kl}n_in_j) 
+ \frac{c}{2}\, (\delta_{ik}n_jn_l+ \delta_{il}n_jn_k + \delta_{jk}n_in_l+ \delta_{jl}n_in_k) 
\nonumber \\ && - \frac13 (a+6b+2c) J_{ijkl} - \frac{2}{15}(a+5c) K_{ijkl} .\quad
\eeqa
The tensors $J_{ijkl}$ and $K_{ijkl}$ are defined in \rf{wr3}, and 
 the elastic constants $a$, $b$ and $c$ follow from \rf{TI1}-\rf{t1} as 
\beq{abc}
a=S_{11}^{(0)} + S_{33}^{(0)} - 2S_{13}^{(0)} - 4S_{44}^{(0)}, \quad 
b=S_{13}^{(0)} - S_{12}^{(0)}, \quad 
c=2S_{44}^{(0)} - 2S_{66}^{(0)}. \quad 
\eeq
  The stress is, as usual, aligned with the fixed axes, so that the total strain energy follows from \rf{fa}, \rf{t1} and \rf{t2}, as 
\beq{t3a}
{\cal E} =  \big[ \frac{1}{\kappa_s} - \frac13 (a+6b+2c)\big] 
\overline{\sigma}^2 + \big[  \frac{1}{2\mu_s} - \frac{2}{15}(a+5c)\big]\sigma_{ij}'\sigma_{ij}' + 
{\cal E}^{\rm (ex)} , 
\eeq
where  $\overline{\sigma}$  is defined in \rf{fis2} and the extra energy term is   
\beq{t4}
{\cal E}^{\rm (ex)} =a\, \big( \sigma_{\rm I}n_1^2+ \sigma_{\rm II}n_2^2+ \sigma_{\rm III}n_3^2\big)^2
+ (3b\overline{\sigma}+ c\sigma_{\rm I})2\sigma_{\rm I}n_1^2
+ (3b\overline{\sigma}+ c\sigma_{\rm II})2\sigma_{\rm II}n_2^2
+ (3b\overline{\sigma}+ c\sigma_{\rm III})2\sigma_{\rm III}n_3^2\, . 
\eeq
The latter shows that the anisotropic part of the energy ${\cal E}^{\rm (ex)}$ depends upon the TI axis orientation through the three parameters 
$n_1^2$, $n_2^2$ and $n_3^2$, which satisfy $n_1^2+n_2^2+n_3^2=1$. Since each of  $n_1^2$, $n_2^2$ and $n_3^2$ must be non-negative, the permissible set is the equilateral  triangular area $A$ of the  plane $n_1^2+n_2^2+n_3^2-1=0$   bounded by the lines $L_1:\, n_2^2+n_3^2=1$, $L_2:\,n_3^2+n_1^2=1$ and $L_3:\,n_1^2+n_2^2=1$.

Consider first the possibility that the optimal orientation lies  
on one of the lines $L_i$, $I=1,2,3$.  Thus, along $L_3$, a simple calculation using $n_3^2 = 0$ shows that 
\beq{t8}
{\cal E}^{\rm (ex)}  = a (\sigma_{\rm II} - \sigma_{\rm I})^2\, \big( n_1^2 - N_1\big)^2 + 2\sigma_{\rm II} \big( 
3b\overline{\sigma} + c \sigma_{\rm II}\big)  - a (\sigma_{\rm II} - \sigma_{\rm I})^2\, N_1^2\, \quad \mbox{on }L_3, 
\eeq
where
\beq{t7}
N_1 \equiv \frac{a \sigma_{\rm II} +c (\sigma_{\rm I} + \sigma_{\rm II}) +3b\overline{\sigma} }{a(\sigma_{\rm II} - \sigma_{\rm I})}\, . 
\eeq
Thus, $n_1^2 = N_1$, is a {\em possible} optimal orientation.  It must first be checked whether or not $N_1$  lies in $(0,1)$. If this is so, and if $a>0$, then an energy minimum occurs at the point $n_1^2=N_1$, $n_2^2=1-N_1$ on $L_3$.  
Similarly, a minimum occurs on $L_1$ at $n_2^2=N_2$, $n_3^2=1-N_2$ if $N_2 \in (0,1)$ and $a>0$, and on $L_2$ at $n_3^2=N_3$, $n_1^2=1-N_3$ if $N_3 \in (0,1)$ and $a>0$, where
\beqa{t9}
N_2 &=& \frac{a \sigma_{\rm III} +c(\sigma_{\rm II} + \sigma_{\rm III}) +3b\overline{\sigma} }{a(\sigma_{\rm III} - \sigma_{\rm II})},
\\\
N_3 &=& \frac{a \sigma_{\rm I} +c(\sigma_{\rm III} + \sigma_{\rm I}) +3b\overline{\sigma} }{a(\sigma_{\rm I} - \sigma_{\rm III})} \, . 
\eeqa 

Now consider the possibility of the minimum occurring in the interior of $A$.   Substitute $n_3^2=1 -n_1^2-n_2^2$ into \rf{t4} and setting the partial derivatives with respect to $n_1^2$ and $n_2^2$ to zero, yields a pair of simultaneous conditions
\beqa{t5}
\big( \sigma_{\rm I} - \sigma_{\rm III}\big)\, \big[ 
a\, \big( \sigma_{\rm I}n_1^2+ \sigma_{\rm II}n_2^2+ \sigma_{\rm III}n_3^2\big)
+ 3b\overline{\sigma}+ c(\sigma_{\rm I}+ \sigma_{\rm III}) \big] &=&0,
\\
\big( \sigma_{\rm II} - \sigma_{\rm III}\big)\, \big[ 
a\, \big( \sigma_{\rm I}n_1^2+ \sigma_{\rm II}n_2^2+ \sigma_{\rm III}n_3^2\big)
+ 3b\overline{\sigma}+ c(\sigma_{\rm II}+\sigma_{\rm III})\big] &=&0 \, . 
\eeqa
These cannot be satisfied in general if the three principle stresses are distinct.   We therefore conclude that \textcolor{blue}{\em the optimal $\bf n$ will lie inside $A$ iff the stress is biaxial}.  
For instance, if $\sigma_{\rm II} $ and $ \sigma_{\rm III}$ are equal, then eq. \rf{t5} combined with $n_3^2=1 -n_1^2-n_2^2$ imply that ${\cal E}^{\rm (ex)}$ of \rf{t4} can be expressed as a function of   $n_1^2$ alone, and the expression is identical in form to that given in \rf{t8}. Thus, the existence of a minimum inside $A$ requires biaxiality $(\sigma_{\rm II}=\sigma_{\rm III})$ and that $0 < N_1 < 1$.  The associated optimal direction is not unique, but is defined by the cone $n_1^2 = N_1$, $n_2^2+n_3^2 = 1-N_1$ (note that $N_3 = 1-N_1$ when $\sigma_{\rm II}=\sigma_{\rm III}$). 
Again, \rf{t8} indicates that the optimal orientation corresponds to a minimum (maximum) in energy if $a>0$ $(a<0)$.   Thus, the sign of the elastic compliance  $a$ is crucial in determining whether the stationary point  is a minimum or a maximum.  

These conclusions may also be confirmed by the coaxiality condition for the stress and strain. Thus, for arbitrary orientation
\beqa{od}
\varepsilon_{23} &=& n_2n_3\, \big[ 
a\big( \sigma_{\rm I}n_1^2+ \sigma_{\rm II}n_2^2+ \sigma_{\rm III}n_3^2\big) 
+ 3b\overline{\sigma}+ c\big( \sigma_{\rm II} + \sigma_{\rm III} \big)\big]\, , 
\nonumber \\
\varepsilon_{31} &=& n_3n_1\, \big[ 
a\big( \sigma_{\rm I}n_1^2+ \sigma_{\rm II}n_2^2+ \sigma_{\rm III}n_3^2\big) 
+ 3b\overline{\sigma}+ c\big( \sigma_{\rm III} + \sigma_{\rm I} \big)\big]\, , 
\\
\varepsilon_{12} &=& n_1n_2\, \big[ 
a\big( \sigma_{\rm I}n_1^2+ \sigma_{\rm II}n_2^2+ \sigma_{\rm III}n_3^2\big) 
+ 3b\overline{\sigma}+ c\big( \sigma_{\rm I} + \sigma_{\rm II} \big)\big]\, . 
\nonumber 
\eeqa
The requirement that these simultaneously  vanish is identical to the  above conditions for the existence of the minimum inside $A$ or along its perimeter.

In summary, $a>0$ is a necessary condition that an energy minimum occurs at points inside $A$ or along the lines $L_j$, $j=1,2,3$.  A minimum is achieved iff one or more of $N_1$, $N_2$ or $N_3$ lies in $(0,1)$.  The minimum occurs on the associated bounding line $L_j$ or on a cone of directions for biaxial states of stress.   Otherwise, the global energy minimum corresponds to one of the  vertices  of $A$, i.e. at $n_1^2 = 1$ or $n_2^2 = 1$ or $n_3^2 = 1$.  In this default case the TI axis of symmetry is aligned with one of the stress axes.  These findings are in agreement with those of Rovati and Taliercio \cite{Rovati03}, who stated the condition as follows: At least one of the principal axes of stress must lie in a plane of transverse isotropy, or alternatively, the TI axis must lie in a plane defined a pair of principal axes of stress. 

\subsection{Tetragonal symmetry} 

The moduli have the same general form as in \rf{TI1}, except that there is no relation between $S_{66}^{(0)}$, $S_{11}^{(0)}$ and $S_{12}^{(0)}$.  In this sense,  tetragonal symmetry is the same as TI but with  one additional elastic constant.  The isotropic moduli are given by eq. \rf{km3}, and the anisotropic part of the compliance is 
\beqa{t1tet}
s_{ijkl}^{\rm (an)} &=& a'\, n_in_jn_kn_l + b'\, (\delta_{ij}n_kn_l+ \delta_{kl}n_in_j) 
+ \frac{c'}{2}\, (\delta_{ik}n_jn_l+ \delta_{il}n_jn_k + \delta_{jk}n_in_l+ \delta_{jl}n_in_k) 
\nonumber \\ && 
+ d\, (p_ip_j- q_iq_j)(p_kp_l - q_kq_l) 
- \frac13 (a'+6b'+2c') J_{ijkl} - \frac{2}{15}(a'+5c' + 3d) K_{ijkl} .\quad \qquad
\eeqa
The additional fourth order tensor as compared to TI is  $ (p_ip_j- q_iq_j)(p_kp_l - q_kq_l)$, 
where $\{ {\bf n}, {\bf p}, {\bf q}\}$ form an orthonormal triad.  
The elastic constants $a'$, $b'$, $c'$  and $d$ are
\beqa{abcd}
&&a'=\frac12 S_{11}^{(0)} +\frac12 S_{12}^{(0)}+ S_{33}^{(0)} + S_{66}^{(0)}  - 2S_{13}^{(0)} - 4S_{44}^{(0)}, 
\nonumber \\ 
&& b'=S_{13}^{(0)} - S_{12}^{(0)} - \frac12 \big( S_{11}^{(0)}- S_{12}^{(0)}\big)  + S_{66}^{(0)}, \quad 
c'=2S_{44}^{(0)} - 2S_{66}^{(0)},  \\
&&
d=\frac12 \big( S_{11}^{(0)}- S_{12}^{(0)}\big)  - S_{66}^{(0)} . \quad \nonumber
\eeqa
Compared to the TI constants $a$, $b$, $c$ of \rf{abc}, 
\beq{abcom}
a'=a-d,\quad b'=b-d,\quad c'=c. 
\eeq

The strain energy of the tetragonal material can be split into a component similar in form to that for a TI material,  and an additional term proportional to the constant $d$.  The minimization of the TI part of the energy is as before (with $a',b',c'$ instead of $a,b,c$), and depends  upon the orientation of $\bf n$ but not $\bf p$ and $\bf q$.   The additional energy term depends  on the deviatoric part of the stress and on these directions, 
\beq{tet2}
{\cal E}^{\rm tet} = d \big[ (p_ip_j-q_iq_j)\sigma_{ij}'\big]^2 - \frac25\, d \sigma_{ij}'\sigma_{ij}', 
\eeq
or, in terms of the principal stresses,  
\beq{tet3}
{\cal E}^{\rm tet} = d\, \big( \sigma_{\rm I} '\Delta_1'  + \sigma_{\rm II} '\Delta_2' + \sigma_3 '\Delta_{\rm III}' \big)^2 - \frac25\, d \sigma_{ij}'\sigma_{ij}' \, , 
\eeq
where
\beq{di}
 \Delta_i'  = p_i^2- q_i^2 \, \quad \mbox{(no sum)} .\quad 
\eeq
The final term in \rf{tet3} is independent of $\{ {\bf n}, {\bf p}, {\bf q}\}$, and it is only  necessary to  consider the quantity
\beq{tq1}
{\cal E} '  = d\, f^2,\quad \mbox{where } 
f = \sigma_{\rm I} '\Delta_1'  + \sigma_{\rm II} '\Delta_2' + \sigma_3 '\Delta_{\rm III}'\, . 
\eeq
The orientation dependence is captured by the quantity $f$.  

It is now demonstrated that for any given $\bf n$, there is at least one set  of $ {\bf p}, {\bf q}$ orthogonal to $\bf n$  which make ${\cal E} '$ vanish.  Let $ {\bf p}^{(0)}, {\bf q}^{(0)}$ be an arbitrary pair of unit vectors such that $\{ {\bf n}, {\bf p}^{(0)}, {\bf q}^{(0)}\}$ form an orthonormal triad, then every possible   set $\{ {\bf n}, {\bf p}, {\bf q}\}$ is defined by the pair $ {\bf p}, {\bf q}$ obtained by rotation about $\bf n$ by angle $\phi$:
\beq{tet5}
{\bf p}(\phi) = \cos \phi \, {\bf p}^{(0)} - \sin \phi \, {\bf q}^{(0)}, 
\qquad 
{\bf q}(\phi) = \sin \phi \, {\bf p}^{(0)} + \cos \phi \, {\bf q}^{(0)}. 
\eeq
It may then be readily verified that 
 \beq{tet6}
\Delta_i'(\phi) = \Delta_i'(0)\, \cos 2\phi - \Delta_i'(\frac{\pi}{4})\, \sin 2\phi , \quad i=1,2,3. 
\eeq
Equation \rf{tet6} implies that 
\beq{tet7}
{\cal E} ' (\phi) = d\, \big[ f(0)\, \cos 2\phi - f(\frac{\pi}{4})\, \sin 2\phi\big]^2 \, , 
\eeq
and hence 
\beq{tet8}
{\cal E} ' (\phi^*) = 0 , \quad \mbox{where } \tan 2\phi^* = f(0)/f(\frac{\pi}{4}) \, , 
\eeq
Thus, if $d>0$, the situation for tetragonal symmetry is a simple addition to the TI situation: First find $\bf n$ which minimizes  the TI part of the energy.  Then, select the pair  $  {\bf p}, {\bf q}$ such that they satisfy \rf{tet8}.  The minimum energy is then exactly that achieved by the TI part of the moduli (although it depends upon $a',b',c'$ rather than $a,b$ and $c$). 

If $d<0$ then the situation is more complicated, and the sequential minimization of first the  TI energy and then  the additional energy ${\cal E} '$ does not work, although these do define stationary points for the strain energy.  The $d-$term must be taken into account when optimizing with respect to $\bf n$, and a more complicated minimization problem is involved. 

Tetragonal symmetry represents a demarcation between the simpler higher material symmetries for which explicit results can be obtained, and the  lower material symmetries which require numerical resolution, in general.  Exceptions may occur, however, it is useful and instructive to distinguish the cubic, TI and tetragonal symmetries from those of, for example, monoclinic symmetry with 13 independent moduli to consider. 

\section{Strain deviation angle}\label{sec5}

\subsection{Definition of the strain deviation angle}

A necessary condition for an energy minimum is that the stress and strain are coaxial.  This is always the case in isotropic media,  whereas it will be the exception rather than the rule under conditions of  general anisotropy and arbitrary stress.  According to Euler's theorem \cite{Baruh} the transformation from one set of principal axes to the other can be reduced to an axis of rotation $\bf n$, $|{\bf n}|=1$, and an angle of rotation $\phi$. The stress axes  have been assumed to coincide with the fixed axes ${\bf e}_j$, $j=1,2,3$.   Let  ${\bf e}_j'$ be the orthonormal axes of the strain tensor, then it follows that the rotation matrix is simply the matrix composed of the three unit vectors as columns, 
\beq{qr}
{\bf Q} = \big[ {\bf e}_1 '\, {\bf e}_2 '\, {\bf e}_3' \big] \, . 
\eeq
Let $\bf Q$ be represented by \rf{qeq}, then it follows from the latter that 
\beq{qr2}
e_{ijk}Q_{jk} = 2 \sin \phi\, n_i.
\eeq
This provides a formula to determine both the angle $\phi$ and the axis of rotation $\bf n$.    

The strain deviation angle $\phi$ is defined as the angle of rotation between the stress and strain axes. This angle is identically zero in isotropic materials for all stress states.  In anisotropic materials it depends on both the material constants and the state of stress.   However,  the above analysis tells us that $\phi = 0$ is a necessary condition for energy minimization. Therefore, the magnitude of $\phi$ provides, through a single parameter, the degree to which the given state of stress and material orientation are optimal.  It does so without requiring any calculation of the energy locally or globally.   It requires only that the principal strain axes are determined, and from those $\phi$  can be immediately computed. 

For a given material, stress and strain, the  strain deviation angle can be obtained from \rf{qr2}.  A more explicit method is to use the  general identity for integer $m$: 
\beq{ged}
  \cos m\phi =  \frac12\, {\rm tr} \big({\bf Q}^m\big) - \frac12. 
\eeq
This follows from, for example, eqs. \rf{qeq} and \rf{a1a1}, which imply
\beq{ged2} {\bf Q}^m  ({\bf n} , \phi ) = {\bf n}\otimes{\bf n} + \cos m\phi \, ({\bf I} -{\bf n}\otimes{\bf n} )+
\sin m\phi\, {\bf P}\, . 
\eeq
For instance, $m=1$ gives the  strain deviation angle explicitly in terms of the first invariant of $\bf Q$: 
\beq{qr3}
  \phi = \cos^{-1}\big[ \frac12 \big({\rm tr}{\bf Q}  -1\big)\big].
\eeq

\subsection{Weak anisotropy}
Let $\varepsilon_{j}^{(0)}$, $j=1,2,3$ be the principal strains for the isotropic medium, i.e. the principal axes of 
$s_{ijkl}^{\rm (is)}\sigma_{kl}$, where $\sigma_{ij}$ is given by \rf{s1} and/or \rf{t2}.  In order to determine the strain deviation angle we first need to find the principal axes of strain.  
It is useful to express the strain as 
\beq{usf}
\varepsilon_{ij} = s_{ijkl}^{\rm (is)}\sigma_{kl}+\gamma_{ij},
\eeq
 where 
\beqa{w0}
\gamma_{ij} &=& s_{ijkl}^{\rm (an)}\sigma_{kl}
\nonumber \\ &=&
s_{ij11}^{\rm (an)}\sigma_{\rm I} + s_{ij22}^{\rm (an)}\sigma_{\rm II}+s_{ij33}^{\rm (an)}\sigma_{\rm III}   
\eeqa
It is assumed that $\bosy{\gamma}$ is small, so that standard perturbation analysis may be used to find the first correction to the directions of principal strain, 
$\{ {\bf e}_1', {\bf e}_2',{\bf e}_3'\}$ , which to leading order are coincident with the stress directions: 
\beq{wa1}
{\bf e}_i' = {\bf e}_i + \sum\limits_{j\ne i}\, \big( \varepsilon_{i}^{(0)} - \varepsilon_{j}^{(0)}\big)^{-1}\, 
\big( {\bf e}_i\cdot \bosy{\gamma}{\bf e}_j\big) \, {\bf e}_j, \, \quad \mbox{no sum on }i. 
\eeq
Let $E$ and $\nu$ be the isotropic Young's modulus and Poisson's ratio characterizing $s_{ijkl}^{\rm (is)}$, then \rf{wa1} implies that, to leading order, 
\beq{wa2}
Q_{ij} = - Q_{ji} \approx \frac{E}{1+\nu}\, \frac{\gamma_{ij}}{ (\sigma_j - \sigma_i)},\quad i\ne j.
\eeq
Hence, the strain deviation angle for weak anisotropy is 
\beq{wa3}
\phi \approx \sin \phi \approx 
 \frac{E}{1+\nu}\, \bigg[ \frac{\gamma_{12}^2}{ (\sigma_{\rm I} - \sigma_{\rm II})^2}
 + \frac{\gamma_{23}^2}{ (\sigma_{\rm II} - \sigma_{\rm III})^2} + \frac{\gamma_{31}^2}{ (\sigma_{\rm III} - \sigma_{\rm I})^2}\bigg]^{1/2}\, . 
\eeq
It is useful to write the stress dependence explicitly by eliminating $\bosy{\gamma}$, 
\beqa{wa31}
\phi &\approx &
 \frac{E}{1+\nu}\, \bigg[  
 \bigg( \frac{\sigma_{\rm I}s_{14}^{\rm (an)}+\sigma_{\rm II}s_{24}^{\rm (an)}+\sigma_{\rm III}s_{34}^{\rm (an)}  }{ \sigma_{\rm II} - \sigma_{\rm III}}\bigg)^2
 +\bigg( \frac{\sigma_{\rm I}s_{15}^{\rm (an)}+\sigma_{\rm II}s_{25}^{\rm (an)}+\sigma_{\rm III}s_{35}^{\rm (an)}  }{ \sigma_{\rm III} - \sigma_{\rm I}}\bigg)^2
 \nonumber \\  && \qquad \qquad
 + \bigg(\frac{\sigma_{\rm I}s_{16}^{\rm (an)}+\sigma_{\rm II}s_{26}^{\rm (an)}+\sigma_{\rm III}s_{36}^{\rm (an)}  }{ \sigma_{\rm I} - \sigma_{\rm II}}\bigg)^2
 \bigg]^{1/2}\, . 
\eeqa
This shows that the strain deviation angle depends upon the same $9$ moduli that appear in the matrix $\bf E$ of eq. \rf{r4}. 

The above formula breaks down for biaxial stress.  In this case the choice of fixed axes is arbitrary since any orthogonal pair in the the plane spanned by the equal principal stresses are valid.  However, the choice can be made {\it a posteriori} such that the term that would otherwise be singular is zero.  For instance, if $\sigma_{\rm I} = \sigma_{\rm II}$, then the axes ${\bf e}_1$ and ${\bf e}_2$ can be selected such that $\gamma_{12}= 0$.  

\section{Conclusions}

The 6-dimensional notation of Mehrabadi et al.  \cite{mcj} is well suited to the problem of finding optimal orientations of anisotropic solids.  It leads quite naturally to the main results of the paper, which we recapitulate:

\medskip 
\noindent
{\bf Result 1}: {The energy  
$ \textcolor{blue} {\cal E}$  is stationary iff the stress and strain are coaxial. }

\noindent
{\bf Result 1a}: {A necessary (but not sufficient) condition for this to hold is that det ${\bf E}=0$, where $\bf E$ is defined in \rf{r4}}

\noindent
{\bf Result 2}:  {The energy
${\cal E}$  is a local minimum if the stress and strain are coaxial and the symmetric matrix ${\bf G}$ of \rf{G7} is positive definite.  } 

Result 2 provides for the first time an explicit set of conditions that must be satisfied if  the stationary condition is a minimum or a maximum.  

Specific  results are given for materials of cubic, transversely isotropic  and tetragonal symmetries.  In each case the existence of a minimum or maximum depends on the sign of a single elastic constant.  For cubic symmetry we have several new findings.  For instance,  eqs. \rf{s3a} - \rf{s3c} provide a two parameter set of stress states which minimize or maximize the strain energy if a material of cubic symmetry is rotated  about an arbitrary axis $\bf n$ by angle $\phi$ (subject to the constraint \rf{n1n2q}).  Alternatively, eqs. \rf{lp} and \rf{exp} provide a means to find the optimal orientation for a given state of stress.   In particular, the rotation of the material axes depends only upon the deviatoric stress.  This demonstrates that the stationary (minimum or maximum) value of energy can always be achieved for cubic materials.  Furthermore, it shows that 
the optimal orientation of a solid with cubic material symmetry is not normally aligned with the symmetry directions.  

The remainder of the new results concern the optimal orientation of  TI and tetragonal materials,   and  are in general agreement with results of  Rovati and Taliercio \cite{Rovati03} obtained by a different procedure.  However, the results obtained here are more direct and provide considerable insight into the nature of the optimal states for these material symmetries.  In particular,  the problem for tetragonal symmetry is very similar to that for TI, with an additional energy term that can be simply minimized or maximized (depending on the sign of the constant $d$ of \rf{abcd}.   

Finally, we have defined and  introduced the strain deviation angle.  The strain deviation angle is inherently anisotropic, and directly related to the problem of energy minimization since the angle   defines the degree to which a state of stress or strain is not optimal.  Future work will explore other consequences of this new concept.

\section*{Appendix: Two dimensional elasticity}

Optimal orientation for  two dimensional elastic anisotropy  is an important special case of the general problem.  It was recently considered by  Gea and Luo \cite{Gea04}, and is reconsidered here  in the context of the present theory.  We will see that some of the features Gea and Luo  obtained transfer to the 3D problem: in particular, the dependence of the minimization upon a single elastic constant. 

The two dimensional strain energy function is 
\beq{q1}
{\cal E}(\theta ) = S_{11} \sigma_{\rm I}^2 + S_{22} \sigma_{\rm II}^2 + 2S_{12} \sigma_{\rm I} \sigma_{\rm II}\, . 
\eeq
where $S_{11}$, $S_{22}$ and $S_{12}$ depend upon the angle $\theta$ by which the material is rotated relative to the fixed $\bf{e}_3$ axis.  Consider an orthotropic material with compliance elements 
$S^{(0)}_{11},\,  S^{(0)}_{22},\,  S^{(0)}_{12},\,  S^{(0)}_{66}$ in the unrotated (fixed) axes.  
Using the standard relations \cite{Christensen} for the transformation of the moduli, it may be shown that 
\beq{q2}
{\cal E}(\theta ) = \frac{1 }{4} \, d_0\, ( \sigma_{\rm II} - \sigma_{\rm I})^2\, \big( 
\cos 2\theta -  \Lambda \big)^2\,  + b_0 ,
\eeq
where $\Lambda$ is a combination of stress and moduli, 
\beq{x6}
 \Lambda  = \bigg( \frac{ \sigma_{\rm II} + \sigma_{\rm I} }{ \sigma_{\rm II} - \sigma_{\rm I} } \bigg) \,
 \frac{ c_0 }{ d_0 }\, , 
 \eeq
 $c_0$ and $d_0$ are   moduli, 
\beq{x1a}
c_0 = S^{(0)}_{11} - S^{(0)}_{22}, \qquad 
d_0 = S^{(0)}_{11} + S^{(0)}_{22} - 2S^{(0)}_{12} - 4S^{(0)}_{66} , 
\eeq
and $b_0$ is a constant,
\beq{b0}
b_0 = ( \sigma_{\rm II} - \sigma_{\rm I})^2\, S^{(0)}_{66} + \frac14 ( \sigma_{\rm II} + \sigma_{\rm I})^2\, \big(
S^{(0)}_{11} + S^{(0)}_{22}+2 S^{(0)}_{12} -\frac{c_0^2}{d_0}\big) . 
\eeq

It can be easily seen that the energy ${\cal E}$ of \rf{q2} is  stationary with respect to $\theta$ when 
\beq{st}
\cos 2\theta = 1, \mbox{ and  } \cos 2\theta = -1
\quad \Leftrightarrow \quad \theta =0, \mbox{ and  } \theta = \pi/2, 
\eeq
respectively.  Which of these yields the  smaller value for ${\cal E}$ depends upon the sign of $d_0\Lambda$, or equivalently, the sign of $ (\sigma_{\rm II}^2 - \sigma_{\rm I}^2)c_0$.  Specifically, the minimum is at $\cos 2\theta = $sgn$[(\sigma_{\rm II}^2 - \sigma_{\rm I}^2)c_0]$. A third stationary value is possible if $-1< \Lambda<1 $, and occurs at 
\beq{ste}
\cos 2\theta = \Lambda, 
\quad \Leftrightarrow \quad \theta  =\pm \theta^*, 
\eeq
where $\theta^* = \frac12\, \cos^{-1}\Lambda$.  If this stationary point occurs, it follows from explicit evaluation that it corresponds to a global minimum or maximum of the energy.  
Thus, 
\beqa{q3}
\left. \ba{c}
{\cal E}(0)\\ \\
{\cal E}(\frac{\pi}{2})\\ \\
{\cal E}(\theta^*)
\ea \right\} = b_0 + d_0\, ( \sigma_{\rm II} - \sigma_{\rm I})^2\, \times \, 
\left\{ \ba{l}
 \sin^4  \theta^* \, , \\ \\
 \cos^4  \theta^* \, , \\ \\
0\, . 
\ea\right. 
\eeqa
It is clear that $\theta = \pm \theta^*$ is a repeated global minimum (maximum) if  $d_0 >0\, (d_0 <0)$. This is the fundamental result of  Gea and Luo \cite{Gea04} (although their conclusion is slightly different since they do not start with the stress in the principal axes frame): 
\textcolor{blue}{ If  $-1< \Lambda<1 $ and $d_0 >0$ then $ \theta  =\pm \theta^*$ is a repeated global minimum of ${\cal E}(\theta)$.  Otherwise, the minimum occurs when $\cos 2\theta = $sgn$[(\sigma_{\rm II}^2 - \sigma_{\rm I}^2)c_0]$.
}

The results of Gea and Luo are now reconsidered within the context of the general theory applied to 2D.   Based on the general theory for 3D, the 2D condition for a stationary orientation is 
\beq{21}
\varepsilon_{12} = 0\, , 
\eeq
or, in terms of the stress, assuming for simplicity that $\sigma_{\rm III} = 0$: 
\beq{22}
S_{16}\sigma_{\rm I} + S_{26}\sigma_{\rm II} = 0\, . 
\eeq
The latter is consistent with the general formulation, i.e. eq. \rf{r3}, under the assumption that $\sigma_{\rm I}$ and $\sigma_{\rm II}$ are distinct (if they are not distinct, then the stress-based energy function is constant for all material orientations). 
The additional condition that the stationary orientation is a local minimum follows from 
\beq{cm8}
\left. \frac{d^2 {\cal E}(\theta)}{d\theta^2 } \right|_{\theta = 0} =  4 F_{33} , 
\eeq
as 
\beq{h3}
F_{33} = 2 (\sigma_{\rm II} - \sigma_{\rm I})^2 \, G_{33} > 0 . 
\eeq
The exact form of $G_{33}$  follows from  \rf{G6}   with $\sigma_{\rm III} = 0$ as  
\beq{h4}
G_{33} = S_{66}  + \frac12 
(\sigma_{\rm II} - \sigma_{\rm I})^{-1}\, \big[ 
	(S_{11}-S_{12})\sigma_{\rm I} + (S_{21}-S_{22})\sigma_{\rm II}  \big]\, . 
\eeq
Rearrangement gives 
\beq{h5}
G_{33} = \frac14 \,
(\sigma_{\rm II} - \sigma_{\rm I})^{-1}\, \big[ (\sigma_{\rm I} + \sigma_{\rm II}) c + 
(\sigma_{\rm I} - \sigma_{\rm II}) d \big]\, , 
\eeq
where $c$ and $d$ are 
\beq{x1}
c = S_{11} - S_{22}, \qquad 
d = S_{11} + S_{22} - 2S_{12} - 4S_{66} . 
\eeq

The specific case of an orthotropic material is considered next.  
It may be shown by use of standard relations \cite{Christensen} that the combinations of moduli in \rf{x1} transform according to 
\beq{x2}
c (\theta) = c_0 \, \cos 2\theta , \qquad 
d (\theta) = d_0 \, \cos 4\theta ,  
\eeq
where 
$c_0 = c(0)$ and $d_0 = d(0)$ are the same moduli defined in \rf{x1a}. 
Also, 
\beqa{x3}
S_{16} &=&-\frac{1}{4} \big( c_0 +  d_0 \cos 2\theta \big) \, \sin 2\theta ,
\\ 
S_{26} &=&-\frac{1}{4} \big( c_0 -  d_0 \cos 2\theta \big) \, \sin 2\theta . 
\eeqa
Hence, 
\beq{x4}
\varepsilon_{12} = -\frac{1}{4}  \sin 2\theta \, 
\big[ (\sigma_{\rm I} + \sigma_{\rm II}) c_0 + (\sigma_{\rm I} - \sigma_{\rm II}) d_0 \cos 2\theta \big]\, . 
\eeq
The strain $\varepsilon_{12}$ vanishes if 
\beq{x5}
 \sin 2\theta = 0,\quad \mbox{or}\quad 
 \cos 2\theta = \Lambda .
 \eeq
Thus, the stationary points are $\theta = 0,\, \pi/2$ and $\pm \theta^*$ where $\cos 2\theta^* = \Lambda $, in agreement with the results of Gea and Luo \cite{Gea04}. 

Using the same notation, \rf{h5} becomes
\beq{qwa}
G_{33} (\theta) =    \frac14 \,d_0\,  \big( \Lambda
\cos 2\theta -  \cos 4\theta \big)\, .
\eeq
In particular, if $- 1<\Lambda <1$, then 
\beq{qw}
G_{33}(\theta^*)  =   \frac14 \, d_0\,  \sin^2 2\theta^*\, .
\eeq
This implies that $\theta= \pm \theta^*$ is a local minimum of ${\cal E}$ iff $d_0$ is positive. The general analysis for 3D optimal orientation does not provide an explicit statement about global minima.   In order to show that it is a global maximum one must compare the value of the energy  ${\cal E}$ at $\theta= \pm \theta^*$ with its value at the other local minimum, as done in \rf{q3}.  

\medskip
\noindent
{\bf Acknowledgment}\,   I would like to acknowledge Hae Chang Gea for useful suggestions.

%%%%%%%%%%%%%%%%%%%    bibliography   %%%%%%%%%%%%%%%%%%%%%%%%%%%%%%%%%%%%%
%						\bibliography{../bib/thermoelastic}
\end{document}